# ABSTRACT SEMICLASSICAL ANALYSIS OF THE VAN HOVE MODEL

## MARCO FALCONI AND LORENZO FRATINI

ABSTRACT. In this paper we study the semiclassical limit $\hbar \to 0$ of a completely solvable model in quantum field theory: the van Hove model, describing a scalar field created and annihilated by an immovable source. Despite its simplicity, the van Hove model possesses many characterizing features of quantum fields, especially in the infrared region. In particular, the existence of non-Fock ground and equilibrium states in the presence of infrared singular sources makes a representation-independent algebraic approach of utmost importance. We make use of recent representation-independent techniques of infinite dimensional semiclassical analysis to establish the Bohr correspondence principle for the dynamics, equilibrium states, and long-time asymptotics in the van Hove model.

The mathematics of quantum fields poses many outstanding challenges, and ranges from algebraic geometry to stochastic PDEs, and it has been subject of active research in mathematics for almost a century [see the monographs BSZ92, DG13, GJ87, Haa92, Spo04, SW00, and references therein for a detailed bibliography and historical recounts]. The so-called *van Hove model* [Van52] is a perfect playground for many crucial aspects of quantum field theory, especially concerning the low momenta (infrared) behavior of massless fields. What makes this model – describing a scalar field whose excitations can be created and destroyed by a fixed source (*e.g.*, a very heavy particle) – so noteworthy is that despite its simplicity (it is exactly solvable) it retains some crucial and perhaps not-so-intuitive features of quantum systems with infinitely many degrees of freedom, allowing to strengthen our intuition – both mathematical and physical – concerning interacting quantum fields. The aim of this paper is to showcase the power and versatility of *abstract semiclassical analysis*, an algebraic version of infinite dimensional semiclassical analysis developed by one of the authors in [Fal18a, Fal18b] building up from the seminal ideas in [AN08], by applying it to the van Hove model and establishing a correspondence principle for such a field theory. The correspondence principle is commonly attributed – not without some historical controversy – to Niels Bohr who formulated it around 1920, even if its version mostly used in contemporary physics was rephrased by Max Born in 1933 as follows (the "new" mechanics being quantum mechanics):

> *It must be demanded that, for the limiting cases of large masses and of orbits of large dimensions, the new mechanics passes over into classical mechanics.*

The correspondence principle is now a staple of quantum theories, and it works as a consistency check of the validity of the mathematical framework of a given quantum theory: at scales in which quantum effects become negligible, classical physics shall emerge from the quantum theory itself. *Semiclassical analysis* is the mathematical framework in which the correspondence principle is naturally formulated. For systems of particles, *i.e.* systems with finitely many degrees of freedom, semiclassical analysis is somewhat rigid (up to isomorphisms, there is a unique classical phase space, a unique topology on it, a unique irreducible representation of the algebra of canonical physical observables, …) and can be formulated very explicitly with the help of Fourier analysis and distribution theory on $\mathbb{R}^{2d}$; in the case of fields, *i.e.* systems with infinitely many degrees of freedom, semiclassical analysis must take into account a much richer structure, most notably the existence of inequivalent representations of the algebra of canonical observables, and benefits from

---

an abstract algebraic formulation. This is done in § 1, while in § 2 the van Hove model is presented, both in its classical and quantum version. The link between the two is established in the form of a correspondence principle in § 3, and constitutes the main novelty of this paper. Exploiting the fact that the van Hove model is explicitly solvable, we are able to prove the correspondence principle quite extensively, for the dynamics (Theorem 3.1), equilibrium states (Theorem 3.2), and long-time asymptotics (scattering theory, Theorem 3.3). Both § 1 and 2 are self-consistent, while § 3 relies heavily on both (especially on § 1.3, 1.4 and 2.3).

***Acknowledgments.*** M.F. acknowledges the supports of PNRR Italia Domani and Next Generation EU through the ICSC National Research Centre for High Performance Computing, Big Data and Quantum Computing; the MUR grant "Dipartimento di Eccellenza 2023-2027" of Dipartimento di Matematica, Politecnico di Milano; and the "Gruppo Nazionale di Fisica Matematica (GNFM)" section of the "Istituto Nazionale di Alta Matematica (INdAM)".

## 1. Abstract semiclassical analysis

In this section we formulate semiclassical analysis from an abstract point of view, tailored for the study of (bosonic) quantum field theories, irrespective of the chosen representation. The presentation is essentially a pedagogic refinement of the basic concepts of [Fal18b], hopefully with a streamlined and simplified notation; it does not amount to much more than an exercise in harmonic and semiclassical analysis. Throughout this section, we use basic notions of category theory (something that is not uncommon for the quantization of fields [see, *e.g.*, Ber66, Nel73]), and make use of the C*-algebraic formulation of quantum (and classical) physics. The reader unfamiliar with the basic concepts of C*-algebras and their applications to physics shall refer to [BR87, BR97, Con94, Tak79, Tak03a, Tak03b]. Starting from the seminal works by Ammari and Nier [AN08, AN09, AN11, AN15], such abstract semiclassical analysis for infinite dimensional systems has been a subject of study in recent years, and it has been successfully applied to several concrete models of physical interest [see ABN19, AF14, AF17, AFH24, AFO23, AFP16, AR20, CCFO21, CFO19, CF18, CFO23a, CFO23b, DFO21, Fal18a, Fal18b, FOR23, LNR15a, LNR15b, LNR16, LNR18, LNR21, and references therein].

### 1.1. The functor of classical observables.

Let us consider a classical field theory, *e.g.* electromagnetism. The energy of the electromagnetic field $(\mathbf{E}, \mathbf{B})$ in a region of space $\Omega$ is given by

$$\int_{\Omega} \big(\mathbf{E}(x)^2 + \mathbf{B}(x)^2\big) \mathrm{d}x \ .$$

It is thus natural to assume that the components of both the electric and magnetic fields are in $L^2(\Omega)$. Also in considering other fields of classical relevance, such as the gravitational field or the density distribution of matter, it is most common to consider them as "rough" functions, or more properly (a subset of) *distributions*. Distributions are defined by duality starting from a space of *test functions*, and test functions play a crucial role in defining the observables of field theory (both classical and quantum). As a matter of fact, for our purposes test functions define uniquely the set of physical observables, in a functorial way.

From the standpoint of classical particle physics, *symplectic structures* come into play only when considering the Hamiltonian dynamics, and there is no need to introduce such a structure in defining observables. However, since it plays a crucial role in the procedure of quantization (see § 1.2 below), it is useful to introduce it already at this stage. In finite dimensional vector spaces, there is only one symplectic form (up to isomorphisms) – the canonical one on $\mathbb{R}^{2d}$ – and it involves the product of canonical observables (the coordinates of $\mathbb{R}^{2d}$, seen as the physical phase space). Multiplying distributions (the canonical field observables) is notoriously a touchy business, while it is typically possible to multiply test functions (think of the Schwartz space $\mathscr{S}(\mathbb{R}^d)$, or compactly supported smooth functions $\mathscr{D}(\mathbb{R}^d)$). Therefore, with quantization in mind, we assume that our space of test functions is endowed with a symplectic structure. This leads to the following definition.

**Definition 1.1** (The space of test functions $\mathscr{T}$)**.**
The *symplectic space of test functions* $(\mathscr{T}, \varsigma)$ is any real vector space $\mathscr{T}$ endowed with a nondegenerate, antisymmetric, bilinear form $\varsigma : \mathscr{T} \times \mathscr{T} \to \mathbb{R}$ (the symplectic form).

Denoting by $\mathbf{Symp}_{\mathbb{R}}$ the (non-small) category of real symplectic vector spaces (with symplectomorphisms as morphisms), we will write with a slight abuse of notation $(\mathscr{T}, \varsigma) \in \mathbf{Symp}_{\mathbb{R}}$ as a given choice of space of test functions.

Given the space of test functions, the *phase space of fields* (and their conjugate momenta) $\mathscr{T}^\star$ shall be given by a suitable space of distributions. There is however some freedom in choosing $\mathscr{T}^\star$, and essentially any vector space in separating duality with $\mathscr{T}$ (*i.e.* such that there exists a bilinear form $B : \mathscr{T} \times \mathscr{T}^\star \to \mathbb{R}$ that separates points in $\mathscr{T}$) works fine. This freedom of choice is in fact fictitious: as we already remarked, for all our purposes it is solely the space of test functions that matters in determining the algebra of classical observables (and thus the fields themselves). Nonetheless, one shall keep most prominently two examples in mind:

- The first, and more general, is to choose $\mathscr{T}^\star = \mathscr{T}^*$, the algebraic dual of linear functionals on $\mathscr{T}$ (perhaps endowed with the weak-* $\sigma(\mathscr{T}^*, \mathscr{T})$ topology, the coarsest topology making all elements of $\mathscr{T}$ continuous when seen as functionals on $\mathscr{T}^*$);

- The second is possible (and natural) whenever $\mathscr{T}$ is also a topological vector space, and it consists in choosing $\mathscr{T}^\star = \mathscr{T}'$ to be the continuous dual of continuous linear functionals on $\mathscr{T}$. This choice agrees with the picture above at the beginning of the section: if we set $\mathscr{S}(\mathbb{R}^d)$ to be the space of test functions, then the classical fields are tempered distributions in $\mathscr{S}'(\mathbb{R}^d)$.

The classical observables are built from $\mathscr{T}$ by considering it the set of characters for the abelian group $\mathscr{T}^\star$: to any $f \in \mathscr{T}$, we associate a character $W_0(f) : \mathscr{T}^\star \to \mathrm{U}(1) = \{z \in \mathbb{C} , |z| = 1\}$ defined by

$$W_0(f) : T \mapsto e^{2\pi i B(f, T)} ,$$

where $B$ is the duality bracket (bilinear form) between $\mathscr{T}$ and $\mathscr{T}^\star$. It is possible to dispense of $\mathscr{T}^\star$ altogether, and see this character group $\{W_0(f) , f \in \mathscr{T}\}$ as an abstract collection of elements in an *abelian C\*-algebra*, obeying to the following rules:

- $W_0(f) \neq 0$ for any $f \in \mathscr{T}$;

- $W_0(f)^* = W_0(-f)$ for any $f \in \mathscr{T}$;

- $W_0(f) W_0(g) = W_0(f + g)$ for any $f, g \in \mathscr{T}$.

These character relations are *C\*-universal* [see BHR04a, BHR04b], hence it is possible to define the C\*-algebra $\mathbb{W}_0(\mathscr{T}, \varsigma)$ *generated* by the characters, *i.e.* the smallest C\*-algebra containing the character group (if a C\*-algebra $\mathfrak{A}$ contains the character group, then $\mathbb{W}_0(\mathscr{T}, \varsigma)$ embeds into $\mathfrak{A}$).

**Lemma 1.2.** *Let* $\mathbb{W}_0(\mathscr{T}, \varsigma) := C^*\{W_0(f) , f \in \mathscr{T}\}$ *be the C\*-algebra generated by the set* $\{W_0(f) , f \in \mathscr{T}\}$, *whose elements obey the following rules:*

- $W_0(f) \neq 0$ *for any* $f \in \mathscr{T}$;

- $W_0(f)^* = W_0(-f)$ *for any* $f \in \mathscr{T}$;

- $W_0(f) W_0(g) = W_0(f + g)$ *for any* $f, g \in \mathscr{T}$;

*and let* $\mathfrak{A}$ *be any C\*-algebra in which* $\mathbb{W}_0(\mathscr{T}, \varsigma)$ *embeds. Then, as elements of* $\mathfrak{A}$, *the characters have the following properties:*

- $W_0(0) = \mathrm{id}_{\mathbb{W}_0(\mathscr{T}, \varsigma)}$ *(the identity element of $\mathbb{W}_0(\mathscr{T}, \varsigma)$);*

- $\|W_0(f)\|_{\mathfrak{A}} = 1$ *for any $f \in \mathscr{T}$;*

- $W_0(f)^* = W_0(f)^{-1}$ *for any $f \in \mathscr{T}$;*

- $\mathrm{spec}(W_0(f)) \subseteq U(1)$ *for any $f \in \mathscr{T}$ (the algebraic spectrum of any elements is contained in the unit circle);*

- $\{W_0(f), f \in \mathscr{T}\}$ *is a unitary representation of the abelian group $(\mathscr{T}, +)$ in $\mathfrak{A}$.*

*Proof.* Let us proceed in an orderly fashion:

- Since $f + 0 = f$ for any $f \in \mathscr{T}$, it follows from the third property of the set $\{W_0(f), f \in \mathscr{T}\}$ that for any $f \in \mathscr{T}$, $W_0(f)W_0(0) = W_0(0)W_0(f) = W_0(f)$. Now, $\{W_0(f), f \in \mathscr{T}\}$ generates $\mathbb{W}_0(\mathscr{T}, \varsigma)$, hence it follows that $W_0(0) = \mathrm{id}_{\mathbb{W}_0(\mathscr{T}, \varsigma)}$.

- We have that $\|W_0(f)\|_{\mathfrak{A}}^2 = \|W_0(f)W_0(f)^*\|_{\mathfrak{A}} = \|W_0(f - f)\|_{\mathfrak{A}} = \|W_0(0)\|_{\mathfrak{A}} = \|\mathrm{id}_{\mathbb{W}_0(\mathscr{T}, \varsigma)}\|_{\mathfrak{A}} = 1$.

- Similarly to the above, $W_0(f)^*W_0(f) = W_0(0) = \mathrm{id}_{\mathbb{W}_0(\mathscr{T}, \varsigma)} = W_0(0) = W_0(f)W_0(f)^*$.

- Since $W_0(f) = W_0(f)^{-1}$ in the C*-algebra $\mathbb{W}_0(\mathscr{T}, \varsigma)$, $W_0(f)$ is *unitary*; it is well-known that the spectrum of unitary elements of a C*-algebra is contained in the complex unit circle.

- $\{W_0(f), f \in \mathscr{T}\}$ is clearly a representation on $\mathfrak{A}$ of $(\mathscr{T}, +)$, and from the above it is also unitary.

$\dashv$

We call the abelian C*-algebra $\mathbb{W}_0(\mathscr{T}, \varsigma)$ generated by the character group $\{W_0(f), f \in \mathscr{T}\}$ the algebra of *almost periodic functions generated by $\mathscr{T}$* (and acting on $\mathscr{T}^\star$) [Boh47], and we call the finite linear combinations

$$\Pi = \sum_{j=1}^{M} \alpha_j W_0(f_j),$$

$\{\alpha_j\}_{j=1}^{M} \subset \mathbb{C}$, the *trigonometric polynomials*. The algebra of almost periodic functions can thus be seen as the uniform norm closure ($\|\Pi\| := \sup_{T \in \mathscr{T}} |\Pi(T)|$) of trigonometric polynomials. Again, let us remark that the symplectic form $\varsigma$ does not play here any role, still we keep it in the notation with quantization in mind, as it will become apparent in § 1.2 below.

To sum up, given a space of test functions $(\mathscr{T}, \varsigma) \in \mathbf{Symp}_{\mathbb{R}}$, we have constructed an algebra of classical (commuting) field observables, the algebra $\mathbb{W}_0(\mathscr{T}, \varsigma)$ of almost periodic functions on the field's phase space $\mathscr{T}^\star$, generated by the test functions $\mathscr{T}$. This construction is *functorial*: let us denote by $\mathbf{C}^*\mathbf{alg}$ the category of C*-algebras, with *-homomorphisms as morphisms, then the map $\mathbb{W}_0$ that associates to each $(\mathscr{T}, \varsigma) \in \mathbf{Symp}_{\mathbb{R}}$ the C*-algebra $\mathbb{W}_0(\mathscr{T}, \varsigma) \in \mathbf{C}^*\mathbf{alg}$ is a functor if it acts on symplectomorphisms $s : (\mathscr{T}, \varsigma) \longrightarrow (\mathscr{U}, \zeta)$ as $\mathbb{W}_0(s) : \mathbb{W}_0(\mathscr{T}, \varsigma) \longrightarrow \mathbb{W}_0(\mathscr{U}, \zeta)$, defined by extending by linearity and density the action on generators

$$\mathbb{W}_0(s)[W_0(f)] = W_0(sf).$$

This leads to the following definition.

**Definition 1.3** (The functor of classical observables)**.**
The functor $\mathbb{W}_0 : \mathbf{Symp}_{\mathbb{R}} \longrightarrow \mathbf{C}^*\mathbf{alg}$, that maps the test functions $\mathscr{T}$ into the algebra of almost periodic functions generated by $\mathscr{T}$ on the field's phase space $\mathscr{T}^\star$, is called the *functor of classical observables*.

Let us conclude this section by briefly discussing the finite dimensional case $\mathscr{T} = \mathbb{R}^{2d}$ (and thus $\mathscr{T}^\star = \mathbb{R}^{2d}$ itself) with the canonical symplectic form $\varsigma_c$, making a parallel with the language used in standard semiclassical analysis. In the latter, the functions on the phase space (classical observables) are called *symbols*, and collections of symbols are called classes [see, *e.g.*, Fol89, Zwo12]. As a class of symbols, the almost periodic functions does not contain, unfortunately, many useful functions, even among the bounded ones. In fact, *no function vanishing at infinity can be almost periodic*. Nonetheless, the structure of regular states built out of the algebra of almost periodic functions is rich enough to develop an interesting semiclassical limit (see § 1.4 below). There are classes of symbols vanishing at infinity that will play a crucial role in what follows, despite not being almost periodic themselves; let us fix a notation for them here. We denote, as above, by $\mathscr{S}'(\mathbb{R}^n)$ the Schwartz space of rapidly decreasing functions, and by $\mathscr{C}_0(\mathbb{R}^n)$ the space of continuous functions vanishing at infinity, endowed with the uniform norm; finally, let us denote by $\mathscr{F}^{-1}(L^1(\mathbb{R}^n))$ the space of functions whose Fourier transform is integrable, that is densely embedded in $\mathscr{C}_0(\mathbb{R}^n)$.

## 1.2. Abstract quantization (by deformation).

To construct quantum observables, the idea is to *deform* classical observables to noncommutative ones. This idea of deformation quantization is very old, and it goes back to P.A.M. Dirac, that first noted that quantum commutators behave, at first order in $i\hbar$, as classical Poisson brackets. Formal and "strict" deformation quantization has become an important tool in studying the algebraic aspects of mathematical physics, and in pure mathematics as well [Gro46, Kon03, Moy49, Rie93, Rie94, see, *e.g.*, the seminal works]. Our take on deformation quantization is different, and it is tailored to the purpose of studying the correspondence principle for fields; we call it *abstract quantization*, and follows closely [Fal18b] (drawing inspiration from [AN08]).

Let us consider the abelian character group $\{W_0(f), f \in \mathscr{T}\}$ defined in § 1.1 above. Since the algebra of classical observables – being that of almost periodic functions – is built upon such abelian group, it seems somewhat natural to take it as a starting point for our deformation. To take into account the experimental observation that at first order in the deformation parameter $\hbar$ (measuring the noncommutativity) the symplectic structure shall come into play through the Poisson bracket (and the fact that in the limit $\hbar \to 0$ the commutativity shall be restored as dictated by the correspondence principle), the simplest choice of deformation of the abelian product is given by

$$(1) \qquad W_h(f)W_h(g) = W_h(f+g)e^{-i\pi^2\hbar\varsigma(f,g)} .$$

The noncommutative character law (1) takes the name of *Weyl relations*. Again, let us consider the noncommutative character group $\{W_h(f), f \in \mathscr{T}\}$ as an abstract collection of elements in a *noncommutative C\*-algebra*, obeying the deformed character group rules:

- $W_h(f) \neq 0$ for any $f \in \mathscr{T}$;

- $W_h(f)^* = W_h(-f)$ for any $f \in \mathscr{T}$;

- $W_h(f)W_h(g) = W_h(f+g)e^{-i\pi^2\hbar\varsigma(f,g)}$ for any $f, g \in \mathscr{T}$.

These elements are commonly called *Weyl operators*; the noncommutative character group they form can be characterized in a fashion that is completely analogous to the abelian character group $\{W_0(f), f \in \mathscr{T}\}$. To do so, let us recall the definition of *Heisenberg group* over $(\mathscr{T}, \varsigma) \in \mathbf{Symp}_{\mathbb{R}}$: the Heisenberg group $\mathbb{H}_h(\mathscr{T}, \varsigma)$ is $\mathscr{T} \times \mathbb{R}$ endowed with the group multiplication

$$(f, \lambda) \cdot (g, \eta) = (f+g, \lambda + \eta - \pi^2\hbar\varsigma(f, g)) .$$

The noncommutative character group is C\*-universal itself [see BHR04a, BHR04b], and thus it is again possible to construct the C\*-algebra $\mathbb{W}_h(\mathscr{T}, \varsigma)$ generated by it.

**Lemma 1.4.** *Let* $\mathbb{W}_h(\mathscr{T}, \varsigma) := C^*\{W_h(f), f \in \mathscr{T}\}$ *be the C\*-algebra generated by the set* $\{W_h(f), f \in \mathscr{T}\}$, *whose elements obey the following rules:*

- $W_h(f) \neq 0$ *for any* $f \in \mathscr{T}$;

- $W_h(f)^* = W_h(-f)$ *for any* $f \in \mathscr{T}$;

- $W_h(f) W_h(g) = W_h(f + g) e^{-i\pi^2 h\varsigma(f,g)}$ *for any* $f, g \in \mathscr{T}$;

*and let* $\mathfrak{A}$ *be any C\*-algebra in which* $\mathbb{W}_h(\mathscr{T}, \varsigma)$ *embeds. Then, as elements of* $\mathfrak{A}$*, the noncommutative characters have the following properties:*

- $W_h(0) = \mathrm{id}_{\mathbb{W}_h(\mathscr{T},\varsigma)}$;

- $\|W_h(f)\|_{\mathfrak{A}} = 1$ *for any* $f \in \mathscr{T}$;

- $W_h(f)^* = W_h(f)^{-1}$ *for any* $f \in \mathscr{T}$;

- $\mathrm{spec}(W_h(f)) \subseteq U(1)$ *for any* $f \in \mathscr{T}$;

- $\{W_h(f), f \in \mathscr{T}\}$ *is a unitary representation of the Heisenberg group* $\mathbb{H}_h(\mathscr{T}, \varsigma)$ *in* $\mathfrak{A}$.

*Proof.* The proof is perfectly analogous to the abelian case, Lemma 1.2. ⊣

The C\*-algebra $\mathbb{W}_h(\mathscr{T}, \varsigma)$ is called the *Weyl C\*-algebra*, or the *C\*-algebra of Canonical Commutation Relations (CCR-algebra)* [see, *e.g.*, BR97, DG13, for a textbook presentation of this C\*-algebra of crucial importance in quantum physics]. The algebra of canonical commutation relations is the natural algebra to describe canonical quantum observables in finite dimensions, and fields with integer spin (*i.e.*, obeying the Bose statistics). Fields with half-integer spin obey the Fermi statistics, and satisfy canonical anti-commutation relations, that among other things imply the Pauli exclusion principle. Due to the exclusion principle, there appear to be some obstructions to the formulation of a satisfactory fermionic Bohr correspondence principle − at least from a mathematical standpoint, where this problem is almost completely open (especially when compared to the deep level of knowledge concerning quantum particles and bosonic fields). In this paper, we limit our attention to bosonic fields.

As in the abelian case, the association of a Weyl C\*-algebra (of "noncommutative almost periodic operators") to a space of test functions is functorial in nature, if one acts on symplectomorphisms through the character group: given $s : (\mathscr{T}, \varsigma) \longrightarrow (\mathscr{U}, \zeta)$, the map $\mathbb{W}_h(s) : \mathbb{W}_h(\mathscr{T}, \varsigma) \longrightarrow \mathbb{W}_h(\mathscr{U}, \zeta)$ is defined through the action on the noncommutative character group

$$\mathbb{W}_h(s)[W_h(f)] = W_h(sf) .$$

This leads to the following definition, essentially dating back to [Seg59, Seg61].

**Definition 1.5** (The Segal functor of quantum observables)**.**
The functor $\mathbb{W}_h : \mathbf{Symp}_{\mathbb{R}} \to \mathbf{C^*alg}$ that maps the test functions $(\mathscr{T}, \varsigma)$ into the Weyl C\*-algebra is called the *Segal functor of quantum observables*.

The complete analogy between the construction of classical and quantum observables starting from test functions is an expression of the *natural quantization procedure*, that we call abstract quantization $\mathfrak{q}_h$. The quantization procedure in this form is very simple and intuitive:

> *Quantization is a map of classical observables (functions in an abelian C\*-algebra) to quantum observables (operators in a nonabelian C\*-algebra).*
>
> *Such a map is obtained by deforming each abelian character, representing unitarily the abelian group* $(\mathscr{T}, +)$ *to a nonabelian character, representing unitarily the Heisenberg group* $\mathbb{H}(\mathscr{T}, \varsigma)$.

The abstract quantization $\mathfrak{q}_h$ is thus defined as a map between the functors $\mathbb{W}_0$ and $\mathbb{W}_h$, and it is "almost" a natural transformation[1]. Given $(\mathscr{T}, \varsigma) \in \mathbf{Symp}_{\mathbb{R}}$, the quantization map $\mathfrak{q}_h$ has component

---

[1]With respect to a natural transformation, it is only densely defined, and its components are not \*-homomorphisms, since they deform the C\*-algebras from abelian to nonabelian (let us call such a natural abelian-to-nonabelian map a natural deformation).

$\mathfrak{q}_h^{\mathscr{T}} : \mathbb{W}_0(\mathscr{T}, \varsigma) \to \mathbb{W}_h(\mathscr{T}, \varsigma)$ densely defined by linearity from its action on the character group:

$$\mathfrak{q}_h^{\mathscr{T}} \big[ W_0(f) \big] = W_h(f) \; .$$

Given a symplectomorphism $s : (\mathscr{T}, \varsigma) \to (\mathscr{U}, \zeta)$, the following diagram is easily checked to be commutative:

$$
\begin{array}{ccc}
\mathbb{W}_0(\mathscr{T}, \varsigma) & \xrightarrow{\;\mathfrak{q}_h^{\mathscr{T}}\;} & \mathbb{W}_h(\mathscr{T}, \varsigma) \\
{\scriptstyle \mathbb{W}_0(s)} \downarrow & & \downarrow {\scriptstyle \mathbb{W}_h(s)} \\
\mathbb{W}_0(\mathscr{U}, \zeta) & \xrightarrow[\;\mathfrak{q}_h^{\mathscr{U}}\;]{} & \mathbb{W}_h(\mathscr{U}, \zeta)
\end{array}
$$

yielding that $\mathfrak{q}_h$ is almost a natural transformation indeed. This leads to the following definition of quantization.

**Definition 1.6** (Abstract quantization).
Abstract quantization is the densely defined natural deformation $\mathfrak{q}_h : \mathbb{W}_0 \to \mathbb{W}_h$ between the functors of classical and quantum observables, that maps each abelian character to its noncommutative deformation.

The components of abstract quantization are all densely defined since trigonometric polynomials are norm-dense in $\mathbb{W}_0(\mathscr{T}, \varsigma)$. Let us remark that it extends naturally to all (continuous or discrete) linear combinations of characters with summable coefficients by Bochner integration: let $X \subseteq \mathscr{T}$, endowed with a $\sigma$-algebra $\Sigma \subseteq 2^X$, and let $\mu$ be a measure on $X$, $g \in \mathbb{C}^X$ be $\mu$-Bochner measurable with respect to the C*-norm (be it on $\mathbb{W}_0$ or $\mathbb{W}_h$); then

$$\mathfrak{q}_h^{\mathscr{T}} \left( \int_X g(x) W_0(x) \mathrm{d}\mu(x) \right) = \int_X g(x) W_h(x) \mathrm{d}\mu(x) \; ,$$

where both integrals shall be intended in the Bochner sense in either $\mathbb{W}_0(\mathscr{T}, \varsigma)$ or $\mathbb{W}_h(\mathscr{T}, \varsigma)$.

To better understand this abstract quantization, let us focus on the finite dimensional case $\mathscr{T} = \mathscr{T}^\star = \mathbb{R}^{2d}$ with the canonical symplectic form $\varsigma_c$. The abstract quantization in this case coincides with the standard *Weyl quantization* $\mathrm{Op}_h^{\mathrm{Weyl}}(a)$ of almost periodic functions [see, *e.g.*, Zwo12]: suppose that $a$ is periodic, and denote by $\hat{a} : \mathbb{Z}^d \to \mathbb{C}$ its Fourier coefficients, *i.e.*

$$a(\,\cdot\,) = \sum_{k \in \mathbb{Z}^d} \hat{a}(k) W_0(k) \; ;$$

then

$$(2) \qquad \mathfrak{q}_h^{\mathbb{R}^{2d}}(a) = \sum_{k \in \mathbb{Z}^d} \hat{a}(k) \, W_h(k) \; .$$

The formula (2) is indeed a common way of writing the standard Weyl quantization of a periodic phase-space symbol $a$ to an operator on $L^2(\mathbb{R}^d)$, by representing the Weyl operator $W_h(k)$ in the (unique irreducible) Schrödinger representation [see DG13, Zwo12, respectively, for a proof of the uniqueness of the Schrödinger irrepresentation and for a detailed analysis of standard Weyl quantization].

The classes of symbols that we will use the most in the following are the ones of cylindrical type on $\mathscr{T}^\star$. Intuitively, a cylindrical function is a function that depends only on the projection of its variable on a suitable set (the base of the cylinder). We provide an alternative definition that fits better our emphasis on the space of test functions $\mathscr{T}$ rather than the phase space of fields $\mathscr{T}^\star$ (the choice of which is somewhat arbitrary, as discussed above). See the discussion below Definition 1.7 for additional details.

**Definition 1.7** (Cylindrical symbol).
A phase-space symbol $a \in \mathbb{W}_0(\mathscr{T}, \varsigma)$ is said to be *cylindrical* if there exists a finite dimensional symplectic subspace of test functions $\mathscr{T} \supseteq F \in \mathbf{Symp}_{\mathbb{R}}$, $\dim F = 2d$, such that there exists a sequence of trigonometric polynomials $\{\Pi_n\}_{n \in \mathbb{N}}$ with all characters from $F$ approximating $a$ in norm: more precisely, there exists

$\{\Pi_n\}_{n\in\mathbb{N}}$ with

$$\Pi_n = \sum_{j=1}^{M_n} \alpha_j(n) W_0(f_j(n))$$

and $\bigcup_{n\in\mathbb{N}}\left\{f_j(n)\,,\, j\in\{1,\dots,M_n\}\right\} \subset F$, such that

$$a = \lim_{n\to\infty}\Pi_n$$

(where the limit is intended in the uniform norm of $\mathbb{W}_0(\mathcal{T},\varsigma)$). The space $F$ is called the *dual base* of the cylindrical symbol.

The terminology "cylindrical" will be recurring from now on, and comes from the following observation. Let us choose $\mathcal{T}^\cdot = \mathcal{T}^*$ as the algebraic dual. Then, there is a natural bijection between any $\sigma(\mathcal{T}^*,\mathcal{T})$-closed subspace $\Phi$ of finite codimension in $\mathcal{T}^*$ and a $\sigma(\mathcal{T},\mathcal{T}^*)$-closed finite dimensional subspace of $\mathcal{T}$: the polar (orthogonal) $\Phi^\circ \subset \mathcal{T}$ of $\Phi$ is in fact a $\sigma(\mathcal{T},\mathcal{T}^*)$-closed finite dimensional subspace of $\mathcal{T}$, and $(\Phi^\circ)^* \cong \mathcal{T}^*/\Phi$ [Bou81]. Therefore, as a function $a : \mathcal{T}^* \to \mathbb{C}$, a cylindrical symbol in $\mathbb{W}_0(\mathcal{T},\varsigma)$ can be seen as a function for which there exists a subspace $\Phi$ of finite codimension ($\Phi = F^\circ$, and called the base of the cylinder) and a function $b : \mathcal{T}^*/\Phi \to \mathbb{C}$ such that

$$a(T) = b(\pi_\Phi T)\,,$$

where $\pi_\Phi : \mathcal{T}^* \to \mathcal{T}^*/\Phi$ is the projection.

**Definition 1.8** (Cylindrical symbol classes).
We denote by $\mathscr{C}_0(\mathcal{T}^\cdot)_{\mathrm{cyl}}$, $\mathscr{S}(\mathcal{T}^\cdot)_{\mathrm{cyl}}$, $\mathscr{C}_c^\infty(\mathcal{T}^\cdot)_{\mathrm{cyl}}$, ... respectively, the space of cylindrical symbols that are continuous and vanishing at infinity/rapidly decreasing/smooth and compactly supported/...

In the light of the above considerations, the quantization formula (2) extends immediately to periodic cylindrical symbols.

Let us conclude this section with the abstract quantization (cylindrical) version of a standard result of semiclassical analysis – the *sharp Gårding inequality*. Moyal brackets, Hörmander symbol classes, and actually the almost totality of the theorems of standard semiclassical analysis can be reformulated (and proved in a purely algebraic way, without resorting to the Schrödinger irrepresentation) for cylindrical symbols and their abstract quantization, without much effort. We leave this exercise to the interested reader, and we state the abstract sharp Gårding inequality without proof [for a proof in standard semiclassical analysis (to be used as a blueprint for the aforementioned exercise) see Zwo12, Theorem 4.32].

**Theorem 1.9** (Abstract sharp Gårding inequality). *Let $a \in \mathscr{C}^\infty(\mathcal{T}^\cdot)_{\mathrm{cyl}}$ periodic symbol with dual base $F \subset \mathcal{T}$, $\dim F = 2d$. If $a \geq 0$, then there exists $C_d \geq 0$ and $\hbar_d > 0$ such that for any $0 < \hbar < \hbar_d$,*

$$(3) \qquad\qquad \mathfrak{q}_\hbar^{\mathcal{T}}(a) \geq -C_d \hbar\,.$$

The inequalities $a \geq 0$ and (3) shall be intended in a C\*-algebraic sense (positivity of the almost periodic function in the abelian case, of the almost periodic operator in the nonabelian case); in particular, it means that the spectrum of $\mathfrak{q}_\hbar^{\mathcal{T}}(a)$ is bounded below by $-C_d\hbar$. We emphasized the dependence of both $C_d$ and $\hbar_d$ on the dimension of the dual base $F$ since such result holds only for the quantization of positive *cylindrical symbols* (the quantization of positive but non-cylindrical symbols, might be unbounded from below).

The abstract quantization does not preserve positivity (even though it preserves boundedness from below); as in finite dimensions, there is another quantization that can be defined from $\mathbb{W}_0(\mathcal{T},\varsigma)$ to $\mathbb{W}_\hbar(\mathcal{T},\varsigma)$ – the so-called *anti-Wick quantization* – that preserves positivity and is defined on the whole algebra. This quantization is less intuitive than $\mathfrak{q}_\hbar$, nonetheless it plays a crucial role in the semiclassical analysis of states (see § 1.3 and 1.4 below). The anti-Wick quantization cannot be defined uniquely on general real symplectic spaces (it depends on a choice of the symplectic basis, and it is thus *unnatural*), but it can be defined uniquely on "Euclidean spaces" (where it is *natural*), *i.e.* on real symplectic spaces that stem from complex inner product spaces. In fact, let $(\mathfrak{h}, \langle \cdot, \cdot \rangle) \in \mathbf{Inn}_{\mathbb{C}}$ be a complex vector space endowed with an inner product (a nondegenerate sesquilinear form); then $(\mathfrak{h}_{\mathbb{R}}, \mathrm{Re}\langle \cdot, \cdot \rangle, \mathrm{Im}\langle \cdot, \cdot \rangle) \in \mathbf{Eucl}_{\mathbb{R}}$ is an "Euclidean space": $\mathfrak{h}_{\mathbb{R}}$ is the real vector space obtained from $\mathfrak{h}$ by forgetting the complex structure (given a vector $\psi \in \mathfrak{h}$, $\psi$ and $i\psi$ are real linear independent in $\mathfrak{h}_{\mathbb{R}}$); $\mathrm{Re}\langle \cdot, \cdot \rangle$ is a real inner product (nondegenerate symmetric bilinear form) on $\mathfrak{h}_{\mathbb{R}}$ and $\mathrm{Im}\langle \cdot, \cdot \rangle$ a symplectic form on $\mathfrak{h}_{\mathbb{R}}$; furthermore, the inner product and symplectic form "agree" as the canonical inner product and symplectic forms in $\mathbb{R}^d$. Below, let us denote by $|\psi| = \sqrt{\langle \psi, \psi \rangle}$ the Euclidean norm on $\mathfrak{h}_{\mathbb{R}}$.

With an abuse of notation, let us still denote by $\mathbb{W}_\hbar$, $\hbar \geq 0$, the functors of classical and quantum observables restricted to Euclidean spaces: given $(\mathfrak{h}, \langle \cdot, \cdot \rangle) \in \mathbf{Inn}_{\mathbb{C}}$, $\mathbb{W}_\hbar(\mathfrak{h}, \langle \cdot, \cdot \rangle) \equiv \mathbb{W}_\hbar(\mathfrak{h}_{\mathbb{R}}, \mathrm{Im}\langle \cdot, \cdot \rangle)$.

**Definition 1.10** (Anti-Wick quantization).

The anti-Wick quantization is the natural deformation $\mathfrak{q}_\hbar^{\mathrm{aw}} : \mathbb{W}_0 \to \mathbb{W}_\hbar$ defined by its components $\mathfrak{q}_\hbar^{\mathfrak{h},\mathrm{aw}} : \mathbb{W}_0(\mathfrak{h}, \langle \cdot, \cdot \rangle) \to \mathbb{W}_0(\mathfrak{h}, \langle \cdot, \cdot \rangle)$, in turn defined by dense extension of the action on trigonometric polynomials $\Pi = \sum_{j=1}^M \alpha_j W_0(f_j)$:

$$\mathfrak{q}_\hbar^{\mathfrak{h},\mathrm{aw}}(\Pi) = \sum_{j=1}^M \alpha_j W_\hbar(f_j) e^{-\frac{\pi^2 \hbar}{2}|f_j|^2} \ .$$

The anti-Wick quantization satisfies the following properties, whose proof is standard and makes use of the so-called FBI transformation [see, *e.g.*, DG13, HMR87, Mar02].

**Proposition 1.11.** *For all $a \in \mathbb{W}_0(\mathfrak{h}, \langle \cdot, \cdot \rangle)$:*

- *$a \geq 0$ implies $\mathfrak{q}_\hbar^{\mathfrak{h},\mathrm{aw}}(a) \geq 0$ ;*

- *$\|\mathfrak{q}_\hbar^{\mathfrak{h},\mathrm{aw}}(a)\| \leq \|a\|$ .*

In the inequality above, the two norms on the left and right hand side are the C*-norms of $\mathbb{W}_\hbar(\mathfrak{h}, \langle \cdot, \cdot \rangle)$ and $\mathbb{W}_0(\mathfrak{h}, \langle \cdot, \cdot \rangle)$, respectively.

### 1.3. States (**classical and quantum**).

The C*-algebraic concept of state is tailored upon the physical one, especially in the nonabelian case. Let us review a special class of states of crucial physical interest in both classical and quantum field theories [Seg61]: the so-called *regular states*.

A (complex) state in a C*-algebra $\mathfrak{A}$ is simply an element $\omega \in \mathfrak{A}'$ of its continuous dual. A *state* is a *positive* element of the dual $\omega \in \mathfrak{A}'_+$ (*i.e.*, $\omega(A) \geq 0$ for any $A \geq 0$), and a *normalized state* $\omega \in \mathfrak{A}'_{+,1}$ is a state of norm one (equivalently, for unital algebras, $\omega(\mathrm{id}) = 1$). Let us start with the abelian algebra $\mathbb{W}_0(\mathcal{T}, \varsigma)$. From a physical perspective, a (normalized) state on classical observables shall be a (probability) measure on the phase space. Let us see that indeed it is the case, if one considers a suitable subset of $\mathbb{W}_0(\mathcal{T}, \varsigma)'$.

**Definition 1.12** (Classical regular states).

A (complex, standard, normalized) state $\omega_0 \in \mathbb{W}_0(\mathcal{T}, \varsigma)'_{(+,1)}$ is *regular* if and only if the $\mathbb{R}$-action

$$\mathbb{R} \ni \lambda \mapsto \omega_0(W_0(\lambda f)) \in \mathbb{C}$$

is continuous for any $f \in \mathcal{T}$.

Let us denote $\mathrm{Reg}_0(\mathcal{T})$ the set of all (complex) regular states, $\mathrm{Reg}_0(\mathcal{T})_+$ the set of regular states, and $\mathrm{Reg}_0(\mathcal{T})_{+,1}$ the set of normalized regular states.

The set of regular states is isomorphic to a set of "measures" on $\mathcal{T}^\star$, called *cylindrical measures*. Given the space of test functions $\mathcal{T}$, the *algebra of cylinders* $C(\mathcal{T}^\star, \mathcal{T})$ on $\mathcal{T}^\star$ is the collection of all cylinder sets[2], that generates the so-called *initial $\sigma$-algebra* $\Sigma(\mathcal{T}^\star, \mathcal{T})$, *i.e.* the smallest $\sigma$-algebra on $\mathcal{T}^\star$ that makes all elements of $\mathcal{T}$ (seen as functionals on $\mathcal{T}^\star$) measurable.

---

[2] A cylinder set $C_{f_1,\dots,f_n}(\mathcal{B})$ is a set in $2^{\mathcal{T}^\star}$ such that there exist $f_1, \dots, f_n \in \mathcal{T}$ and a Borel set $\mathcal{B} \subseteq \mathbb{R}^n$ such that $C_{f_1,\dots,f_n}(B) = \{T \in \mathcal{T}^\star , \ (B(f_1, T), \dots, B(f_n, T)) \in \mathcal{B}\}$.

**Definition 1.13** (Cylindrical measures)**.**
A map $\mathfrak{m} : C(\mathcal{T}^\star, \mathcal{T}) \to \mathbb{R}_+$ is called a *cylindrical measure* if and only if $\mathfrak{m}$ is $\sigma$-additive when restricted to $C(\mathcal{T}^\star, F) = \Sigma(\mathcal{T}^\star, F)$, for any subspace $F \subseteq \mathcal{T}$ of finite dimension.

Let us call $M = \mathfrak{m}(\mathcal{T}^\star)$ the *total mass* of the cylindrical measure $\mathfrak{m}$, and let us denote by $\mathcal{M}(\mathcal{T}^\star)_{\text{cyl}}$ the space of cylindrical measures. Furthermore, let us denote by $\mathcal{P}(\mathcal{T}^\star)_{\text{cyl}} \subset \mathcal{M}(\mathcal{T}^\star)_{\text{cyl}}$ the set of *cylindrical probabilities* with total mass one, and by $\mathcal{M}_{\mathbb{C}}(\mathcal{T}^\star)_{\text{cyl}}$ the space of complex cylindrical measures.

Cylindrical measures are uniquely identified by their Fourier transform: let us consider the character $W_0(f) = e^{2\pi i B(f, \cdot)}$. Since it depends only on $f \in \mathcal{T}$, $W_0(f)$ is $\Sigma(\mathcal{T}^\star, \text{span}\{f\})$-measurable, and $\mathfrak{m} \in \mathcal{M}(\mathcal{T}^\star)_{\text{cyl}}$ is a $\sigma$-additive finite measure on that $\sigma$-algebra. Therefore, $W_0(f)$ is $\mathfrak{m}$-integrable, and define

$$\hat{\mathfrak{m}}(f) = \int_{\mathcal{T}^\star}^{\bullet} W_0(f)[T] \, \mathrm{d}\mathfrak{m}(T)$$

the *Fourier transform* $\hat{\mathfrak{m}}$ of $\mathfrak{m}$, evaluated in $f$. The notation $\int_{\mathcal{T}^\star}^{\bullet} (\,\cdot\,) \, \mathrm{d}\mathfrak{m}(T)$ highlights the fact that it is only possible to integrate $\Sigma(\mathcal{T}^\star, F)$-measurable functions with respect to a cylindrical measure ($F$ being finite dimensional). The well-known Bochner theorem extends to cylindrical measures.

**Theorem 1.14** (Bochner theorem for cylindrical measures [VTC87])**.** *The Fourier transform is a bijection between* $\{\mathfrak{m} \in \mathcal{M}(\mathcal{T}^\star)_{\text{cyl}}, \; \mathfrak{m}(\mathcal{T}^\star) = M\}$ *and the subset* $\mathfrak{G}(M) \subset \mathbb{C}^{\mathcal{T}}$ *of complex-valued functions on $\mathcal{T}$ that satisfy: for any* $G \in \mathfrak{G}(M)$,

- $G(0) = M$

- $G$ *is positive definite: for any* $n \in \mathbb{N}_*$, $\{\alpha_j\}_{j=1}^n \subset \mathbb{C}$, $\{f_j\}_{j=1}^n \subset \mathcal{T}$,

$$\sum_{j,k=1}^{n} \bar{\alpha}_k \alpha_j G(f_j - f_k) \geq 0$$

- $G\big|_F$ *is continuous for any subspace $F \subset \mathcal{T}$ of finite dimension.*

Thanks to the Bochner theorem, we can prove that there is a bijection between regular states on $\mathbb{W}_0(\mathcal{T}, \varsigma)$, and cylindrical measures (that are uniquely identified by $\mathcal{T}$, as it is apparent from Theorem 1.14).

**Proposition 1.15.** *There are bijections between:*

- $\text{Reg}_0(\mathcal{T})$ *and* $\mathcal{M}_{\mathbb{C}}(\mathcal{T}^\star)_{\text{cyl}}$;

- $\text{Reg}_0(\mathcal{T})_+$ *and* $\mathcal{M}(\mathcal{T}^\star)_{\text{cyl}}$;

- $\text{Reg}_0(\mathcal{T})_{+,1}$ *and* $\mathcal{P}(\mathcal{T}^\star)_{\text{cyl}}$.

*Proof.* Let us provide the proof for the bijection $\text{Reg}_0(\mathcal{T})_{+,M}$ of regular states of norm $M$ and $\{\mathfrak{m} \in \mathcal{M}(\mathcal{T}^\star)_{\text{cyl}}, \; \mathfrak{m}(\mathcal{T}^\star) = M\}$, from which the last two bijections follow immediately, and the first follows from the unique decomposition of a complex state in four positive states in a C*-algebra.

Let us first prove that $\text{Reg}_0(\mathcal{T})_{+,M}$ injects into $\{\mathfrak{m} \in \mathcal{M}(\mathcal{T}^\star)_{\text{cyl}}, \; \mathfrak{m}(\mathcal{T}^\star) = M\}$. Let $\omega_0 \in \text{Reg}_0(\mathcal{T})_{+,M}$, and define $G_{\omega_0} \in \mathbb{C}^{\mathcal{T}}$ by

$$G_{\omega_0}(f) := \omega_0(W_0(f)) \,.$$

It follows that $G_{\omega_0} \in \mathfrak{G}(M)$:

- $G_{\omega_0}(0) = \omega_0(W_0(0)) = \omega_0(\text{id}) = M$;

- $\sum_{j,k} \bar{\alpha}_k \alpha_j G_{\omega_0}(f_j - f_k) = \omega_0\big((\sum_k \alpha_k W_0(f_k))^* (\sum_j \alpha_j W_0(f_j))\big) \geq 0$, since $\omega_0$ is positive (and $A^*A \geq 0$ for any element of a C*-algebra);

- $G_{\omega_0}\big|_F$ is continuous for any finite dimensional subspace $F \subset \mathscr{T}$ by the regularity condition that all $\mathbb{R}$-actions $\lambda \mapsto \omega_0(W_0(\lambda f))$ are continuous.

Then by Bochner's theorem, since the expectation of the abelian character group defines uniquely a state on the algebra, $\mathrm{Reg}_0(\mathscr{T})_{+,M}$ injects into $\{\mathfrak{m} \in \mathscr{M}(\mathscr{T}^\star)_{\mathrm{cyl}}, \ \mathfrak{m}(\mathscr{T}^\star) = M\}$.

It remains to prove that the map is surjective, *i.e.* that given any cylindrical measure $\mathfrak{m}$ with mass $M$, it defines a state on $\mathbb{W}_0(\mathscr{T}, \varsigma)$. Let $\Pi = \sum_{j=1}^n \alpha_j W_0(f_j)$ be a trigonometric polynomial. By the $\sigma$-additivity of $\mathfrak{m}$ on $\Sigma(\mathscr{T}^\star, \mathrm{span}\{f_j\}_{j=1}^n)$, it follows that $\Pi$ is $\mathfrak{m}$-integrable, and define

$$\omega_{0,\mathfrak{m}}(\Pi) := \int_{\mathscr{T}^\star}^{\bullet} \Pi(T)\mathrm{d}\mathfrak{m}(T) = \sum_{j=1}^n \alpha_j \hat{\mathfrak{m}}(f_j) \ .$$

By density of the trigonometric polynomials in $\mathbb{W}_0(\mathscr{T}, \varsigma)$, $\omega_{0,\mathfrak{m}}$ extends uniquely to a positive bounded linear functional on $\mathbb{W}_0(\mathscr{T}, \varsigma)$, *i.e.* to a regular state. Its norm is given by

$$\omega_{0,\mathfrak{m}}(\mathrm{id}) = \omega_{0,\mathfrak{m}}(W_0(0)) = \hat{\mathfrak{m}}(0) = M \ .$$

⊣

So the regular states on the algebra of almost periodic functions are the cylindrical measures. Cylindrical measures (and thus regular states) can also be seen as the dual of $\mathscr{C}_0(\mathscr{T}^\star)_{\mathrm{cyl}}$, by extending the Riesz-Markov theorem to the cylindrical case. The proof of this result is again left as an exercise to the interested reader, and it amounts to gluing up together the standard Riesz-Markov theorem on the finite dimensional subspaces of $\mathscr{T}$.

**Theorem 1.16** (Cylindrical Riesz-Markov theorem). *There are bijections between:*

- $(\mathscr{C}_0(\mathscr{T}^\star)_{\mathrm{cyl}})'$ *and* $\mathscr{M}_{\mathbb{C}}(\mathscr{T}^\star)_{\mathrm{cyl}}$ ;

- $(\mathscr{C}_0(\mathscr{T}^\star)_{\mathrm{cyl}})'_+$ *and* $\mathscr{M}(\mathscr{T}^\star)_{\mathrm{cyl}}$ ;

- $(\mathscr{C}_0(\mathscr{T}^\star)_{\mathrm{cyl}})'_{+,1}$ *and* $\mathscr{P}(\mathscr{T}^\star)_{\mathrm{cyl}}$ .

Since by using abstract quantization $\mathfrak{q}_h$ we pass (almost) naturally from $\mathbb{W}_0(\mathscr{T}, \varsigma)$ to $\mathbb{W}_h(\mathscr{T}, \varsigma)$, it shall come to no surprise that the above characterization works, *mutatis mutandis* at the quantum level by substituting the abelian character group with the noncommutative character group.

**Definition 1.17** (Quantum regular states).
A (complex, standard, normalized) state $\omega_h \in \mathbb{W}_h(\mathscr{T}, \varsigma)'_{(+,1)}$ is *regular* if and only if the $\mathbb{R}$-action

$$\mathbb{R} \ni \lambda \mapsto \omega_h(W_h(\lambda f)) \in \mathbb{C}$$

is continuous for any $f \in \mathscr{T}$.

Let us denote $\mathrm{Reg}_h(\mathscr{T}, \varsigma)$ the set of all (complex) regular states, $\mathrm{Reg}_h(\mathscr{T}, \varsigma)_+$ the set of regular states, and $\mathrm{Reg}_h(\mathscr{T}, \varsigma)_{+,1}$ the set of normalized regular states.

Quantum regular states are well-studied in the literature [see, *e.g* BSZ92, BR97, Seg59, Seg61], and have other equivalent definitions to the one above. In particular, quantum regular states can be represented as density matrices in the (unique irreducible) Schrödinger representation when reduced to the subalgebras $\mathbb{W}_h(F, \varsigma|_F)$, for any $F \subset \mathscr{T}$ symplectic subspace of finite dimension. Quantum regular states can also be characterized as *noncommutative cylindrical measures*, and in fact they satisfy a noncommutative Bochner theorem. Let us define the *noncommutative Fourier transform* of a quantum state $\omega_h$ as the map $\hat{\omega}_h : \mathscr{T} \to \mathbb{C}$

defined by

$$\tilde{\omega}_\hbar(f) := \omega_\hbar(W_\hbar(f)) \, .$$

Let us denote the dual norm of a state $\omega_\hbar \in \mathbb{W}_\hbar(\mathscr{T}, \varsigma)'$ simply by $\|\omega_\hbar\| = \omega_\hbar(W_\hbar(0))$, if no confusion arises.

**Theorem 1.18** (Noncommutative Bochner theorem [Seg61]). *The noncommutative Fourier transform is a bijection between $\{\omega_\hbar \in \mathrm{Reg}_\hbar(\mathscr{T}, \varsigma)_+ \, , \, \|\omega_\hbar\| = M\}$ and the subset $\mathfrak{G}_\hbar(M) \subset \mathbb{C}^{\mathscr{T}}$ of complex-valued functions on $\mathscr{T}$ that satisfy: for any $G_\hbar \in \mathfrak{G}_\hbar(M)$,*

- $G_\hbar(0) = M$

- $G_\hbar$ *is quantum positive definite: for any* $n \in \mathbb{N}_*$, $\{\alpha_j\}_{j=1}^n \subset \mathbb{C}$, $\{f_j\}_{j=1}^n \subset \mathscr{T}$,

$$\sum_{j,k=1}^n \bar{\alpha}_k \alpha_j G_\hbar(f_j - f_k) e^{-i\pi^2 \hbar \varsigma(f_j, f_k)} \geq 0$$

- $G_\hbar\big|_F$ *is continuous for any subspace $F \subset \mathscr{T}$ of finite dimension.*

The only difference between the Fourier transform of a cylindrical measure and the noncommutative Fourier transform of a noncommutative cylindrical measure is the notion of positive definiteness: in the quantum case, positive definiteness holds up to a $\hbar$-dependent deformation, that reflects the deformation leading to the noncommutativity of quantum characters.

## 1.4. Abstract Wigner measures.

In this section we connect commutative and noncommutative cylindrical measures through the classical limit $\hbar \to 0$ (and the action of $\mathfrak{q}_\hbar^{\mathscr{T}}$). In the finite dimensional case, such connection is well-established, and the semiclassical measures obtained starting from a quantum state are called *Wigner measures* [see LP93, of which we follow closely the proof strategy as well].

The semiclassical limit $\hbar \to 0$ shall be seen intuitively as a sort of "adjoint" to quantization, linking back quantum observables to classical ones, as well as quantum to classical states. There are, however, many caveats, as we try to emphasize below.

Firstly, let us consider $a \in \mathscr{F}^{-1}L^1(\mathscr{T}^\star)_{\mathrm{cyl}} \subset \mathscr{C}_0(\mathscr{T}^\star)_{\mathrm{cyl}}$ with dual base $F$. As discussed above, such symbol is not almost periodic, since it vanishes at infinity. Nonetheless, in the GNS representation $(\mathscr{H}_{\omega_\hbar}, \pi_{\omega_\hbar}, \Omega_{\omega_\hbar})$ of any (regular) state $\omega_\hbar$ [see, e.g. BR87, Tak79], the function $\kappa \mapsto \hat{a}(\kappa)\pi_{\omega_\hbar}(W_\hbar(\kappa))$ is Lebesgue Bochner integrable with respect to the operator norm on $\mathscr{H}_{\omega_\hbar}$, and thus

$$\int_F \hat{a}(\kappa)\pi_{\omega_\hbar}(W_\hbar(\kappa)) \mathrm{d}\kappa$$

defines a bounded operator on $\mathscr{H}_{\omega_\hbar}$, and the same is true for

$$\int_F \hat{a}(\kappa)\pi_{\omega_\hbar}(W_\hbar(\kappa)) e^{-\frac{\pi^2\hbar}{2}|\kappa|^2} \mathrm{d}\kappa \, ,$$

supposing $\mathscr{T} = \mathfrak{h}_{\mathbb{R}}$, $\mathfrak{h}$ a complex inner product space. Therefore, with an abuse of notation, we can define

$$\mathfrak{q}_\hbar^{\mathscr{T}}(a) \, , \, \mathfrak{q}_\hbar^{\mathfrak{h},\mathrm{aw}}(a)$$

for any $a \in \mathscr{F}^{-1}L^1(\mathscr{T}^\star)_{\mathrm{cyl}}$.

**Theorem 1.19** (Existence of abstract Wigner measures). *Let $\{\omega_\hbar \in \mathrm{Reg}_\hbar(\mathscr{T}, \varsigma)_+ \, , \, \hbar \in (0, \underline{\hbar})\}$ be a bounded family of quantum regular states:*

$$\sup_{\hbar \in (0, \underline{\hbar})} \|\omega_\hbar\| = M < +\infty \, .$$

*Then there exists a net $\hbar_\beta \to 0$ and a cylindrical measure $\mathfrak{m} \in \{\mathfrak{n} \in \mathcal{M}(\mathcal{T}^\bullet)_{\mathrm{cyl}} , \ \mathfrak{n}(\mathcal{T}^\bullet) \leq M\}$ such that for any $a \in \mathscr{F}^{-1}L^1(\mathcal{T}^\bullet)_{\mathrm{cyl}}$,*

$$\lim_{\hbar_\beta \to 0} \omega_{\hbar_\beta}(\mathfrak{q}_{\hbar_\beta}^{\mathcal{T}}(a)) = \int_{\mathcal{T}^\bullet}^\bullet a(T)\mathrm{d}\mathfrak{m}(T) . \tag{4}$$

*Any measure $\mathfrak{m}$ such that there exists a net $\hbar_\beta \to 0$ such that (4) holds is called a Wigner measure. Given a bounded family of regular states $\{\omega_\hbar, \hbar \in (0, \underline{\hbar})\}$, let us denote by $\mathrm{Wig}\{\omega_\hbar, \hbar \in (0, \underline{\hbar})\}$ the (nonempty) set of all its Wigner measures.*

*Remark* 1.20. A few comments are in order:

- The Wigner measure might lose mass with respect to the quantum state: even if we take normalized quantum states, their Wigner measures can have any mass between 0 and 1! This effect, also well known and studied in finite dimensions, is due to the fact that part or all of the mass of the state might concentrate "at infinity" in the phase space (at least intuitively, the picture shall be clear if one thinks to the case in which $\mathcal{T}$ is endowed with a Fréchet topology).

- The convergence of observables is restricted to the quantization of symbols that vanish at infinity. Therefore, in general we are not guaranteed that the noncommutative Fourier transform, or the expectation of quantized trigonometric polynomials converge to their classical counterparts. This is due to the fact that these functions are supported "at infinity" in the phase space, so they are affected by the eventual loss of mass discussed above.

- There can be more than a single Wigner measure associated to a family of states, as the existence of a cluster point follows from a compactness argument, as shown in the proof of the theorem given below. There are a few instances, for example when looking at quantum ergodicity [see, *e.g.*, the seminal work Col85], that the existence of more than one cluster point is of physical relevance.

- We kept the terminology that is mostly used in the semiclassical literature. Of course, in view of Proposition 1.15, the Wigner measures are also regular states in $\mathrm{Reg}_0(\mathcal{T})_+$, and thus the convergence can be reformulated as a convergence of regular quantum to classical states. In what follows, we will use the notations $\mathfrak{m}$ and $\omega_0$ for cylindrical measures/regular classical states interchangeably (but, hopefully, in a consistent way).

*Proof of Theorem 1.19.* For the sake of simplicity, let us restrict to Euclidean spaces stemming from $(\mathfrak{h}, \langle \cdot, \cdot \rangle) \in \mathbf{Inn}_{\mathbb{C}}$, the general case being obtained exploiting any anti-Wick quantization on $(\mathcal{T}, \varsigma) \in \mathbf{Symp}_{\mathbb{R}}$ by fixing a symplectic basis.

Let us consider the space of cylindrical symbols $\mathscr{F}^{-1}L^1(\mathfrak{h}^\bullet)_{\mathrm{cyl}} =: \mathscr{L}$. We can endow $\mathscr{L}$ with a norm $\|\cdot\|_1$, defined as

$$\|a\|_1 = \int_F |\hat{a}(k)|\mathrm{d}k ,$$

where $F$ is the dual base of $a$. Its continuous dual $\mathscr{L}'$ is also a normed space when endowed with the dual norm $\|\omega\|_{\mathscr{L}'} = \sup_{\|a\|_1 = 1} |\omega(a)|$. By the Banach-Alaoglu theorem, the set

$$L(M) = \{\omega \in \mathscr{L}' , \ \|\omega\|_{\mathscr{L}'} \leq M\}$$

is compact in the weak $\sigma(\mathscr{L}', \mathscr{L})$ topology.

Now, given any regular state $\omega_\hbar \in \mathrm{Reg}_\hbar(\mathfrak{h}, \langle \cdot, \cdot \rangle)_+$ of the semiclassical family, denote by $M_\hbar$ its mass $(\omega_\hbar(\mathbb{1}) = M_\hbar)$; $\omega_\hbar$ defines a continuous linear functional $G_{\omega_\hbar} : \mathscr{L} \to \mathbb{C}$ by

$$G_{\omega_\hbar}(a) = \omega_\hbar(\mathfrak{q}_\hbar^{\mathfrak{h}}(a)) = \langle \Omega_{\omega_\hbar}, \left(\int_F \hat{a}(k)\pi_{\omega_\hbar}(W_\hbar(k))\mathrm{d}k\right)\Omega_{\omega_\hbar}\rangle_{\mathscr{H}_{\omega_\hbar}} .$$

Linearity is easy to check, while continuity follows from

$$\left|G_{\omega_\hbar}(a)\right| \leq \int_F \left|\hat{a}(k)\right| \left|\omega_\hbar(W_\hbar(k))\right| \mathrm{d}k \leq M_\hbar \|\hat{a}\|_{L^1} = M_\hbar \|a\|_1 \ .$$

It follows that $\{G_{\omega_\hbar} \ , \ \hbar \in I\} \subset L(M)$ is a relatively compact set in $\mathscr{L}'$ with respect to the $\sigma(\mathscr{L}', \mathscr{L})$ topology. Hence there exists a net $\hbar_\alpha \to 0$ and $\omega_0 \in L(M)$ such that

$$\lim_{\hbar_\alpha \to 0} G_{\omega_{\hbar_\alpha}} = \omega_0$$

in the $\sigma(\mathscr{L}', \mathscr{L})$ topology; in particular, for any $a \in \mathscr{L}$,

$$\lim_{\hbar_\alpha \to 0} \omega_{\hbar_\alpha}(\mathfrak{q}_{\hbar_\alpha}^\flat(a)) = \lim_{\hbar_\alpha \to 0} G_{\omega_{\hbar_\alpha}}(a) = \omega_0(a) \ .$$

It remains to show that $\omega_0 \in \mathrm{Reg}_0(\mathfrak{h}, \langle \cdot, \cdot \rangle)_+$, *i.e.* it is a classical regular state. Its positivity follows from the Gårding inequality – Theorem 1.9 – however we prove it by other means together with the fact that $\omega_0$ extends uniquely to a continuous linear functional on $\mathbb{W}_0(\mathfrak{h}, \langle \cdot, \cdot \rangle)$.

The semiclassical family of states also defines an element of the dual of $\mathscr{C} := \mathscr{C}_0(\mathfrak{h}^\star)_{\mathrm{cyl}}$, endowed with the uniform norm. Now define $H_{\omega_\hbar} \in \mathscr{C}'$ by uniquely extending the densely defined functional

$$H_{\omega_\hbar}(a) = \omega_\hbar(\mathfrak{q}_\hbar^{\flat,\mathrm{aw}}(a)) = \langle \Omega_{\omega_\hbar}, \left(\int_F \hat{a}(k)\pi_{\omega_\hbar}(W_\hbar(k))e^{-\frac{\pi^2\hbar}{2}|k|^2}\mathrm{d}k\right)\Omega_{\omega_\hbar}\rangle_{\mathscr{H}_{\omega_\hbar}} \ ,$$

for any $a \in \mathscr{L}$ ($\mathscr{L}$ is dense in $\mathscr{C}$ with respect to the topology induced by the uniform norm). The functional is clearly linear, and its continuity follows from Proposition 1.11:

$$\left|H_{\omega_\hbar}(a)\right| \leq M_\hbar \|a\| \ .$$

Furthermore, again by Proposition 1.11, $H_{\omega_\hbar} \in \mathscr{C}'_+$ is *positive*. It then follows, again by the Banach-Alaoglu theorem, that $\{H_{\omega_\hbar} \ , \ \hbar \in I\}$ is relatively compact in $\mathscr{C}'_+$ with respect to the $\sigma(\mathscr{C}'_+, \mathscr{C})$ topology. Therefore, given the net $\hbar_\alpha \to 0$ above that guaranteed convergence of $G_{\omega_{\hbar_\alpha}}$, there exists a subnet $\hbar_{\alpha_\beta} \to 0$ and $\varpi_0 \in \mathscr{C}'_+$, $0 \leq \|\varpi_0\|_{\mathscr{C}'} \leq M$ such that for all $a \in \mathscr{L}$,

$$\lim_{\hbar_{\alpha_\beta} \to 0} \omega_{\hbar_{\alpha_\beta}}(\mathfrak{q}_{\hbar_{\alpha_\beta}}^{\flat,\mathrm{aw}}(a)) = \lim_{\hbar_{\alpha_\beta} \to 0} H_{\omega_{\hbar_{\alpha_\beta}}}(a) = \varpi_0(a) \ .$$

By the cylindrical Riesz-Markov theorem, Theorem 1.16, $\varpi_0 \in \mathrm{Reg}_0(\mathscr{T}, \varsigma)_+$ is a regular state, and $0 \leq \varpi_0(\mathbb{1}) = \|\varpi_0\|_{\mathscr{C}'} \leq M$.

Let now $\hbar_{\alpha_\gamma} \to 0$ be any subnet of $\hbar_\alpha \to 0$ such that there exists $\eta_0 \in \mathrm{Reg}_0(\mathscr{T}, \varsigma)_+$ yielding, for any $a \in \mathscr{L}$,

$$\lim_{\hbar_{\alpha_\gamma} \to 0} \omega_{\hbar_{\alpha_\gamma}}(\mathfrak{q}_{\hbar_{\alpha_\gamma}}^{\flat,\mathrm{aw}}(a)) = \lim_{\hbar_{\alpha_\gamma} \to 0} H_{\omega_{\hbar_{\alpha_\gamma}}}(a) = \eta_0(a) \ .$$

Then for any $a \in \mathscr{L}$, by dominated convergence theorem,

$$|\eta_0(a) - \omega_0(a)| = \lim_{\hbar_{\alpha_\gamma} \to 0} \left|\omega_{\hbar_{\alpha_\gamma}}(\mathfrak{q}_{\hbar_{\alpha_\gamma}}^{\flat,\mathrm{aw}}(a) - \mathfrak{q}_{\hbar_{\alpha_\gamma}}^\flat(a))\right| \leq M \lim_{\hbar_{\alpha_\gamma} \to 0} \int_F |\hat{a}(k)||e^{-\frac{\pi^2\hbar}{2}|k|^2} - 1|\mathrm{d}k = 0 \ .$$

It follows that $\eta_0|_{\mathscr{L}} = \omega_0$; however since for any finite dimensional subspace $F \subseteq \mathfrak{h}$, $\mathscr{F}^{-1}L^1(\mathfrak{h}^\star/F^\circ)$ is dense in $C_0(\mathfrak{h}^\star/F^\circ)$, $\mathscr{L}$ separates points in $\mathscr{C}'$. Therefore, $\omega_0$ extends uniquely to a positive continuous linear functional on $\mathscr{C}'$, and thus by the cylindrical Riesz-Markov theorem to a regular state $\omega_0 \in \mathrm{Reg}_0(\mathscr{T}, \varsigma)_+$ with $0 \leq \omega_0(\mathbb{1}) \leq M$. Finally, by uniqueness of the cluster point of $\{\omega_{\hbar_\alpha} \ , \ \hbar_\alpha \to 0\}$ in $\mathscr{C}'_+$, it also holds true that for any $a \in \mathscr{L}$,

$$\lim_{\hbar_\alpha \to 0} \omega_{\hbar_\alpha}(\mathfrak{q}_{\hbar_{\alpha_\gamma}}^{\flat,\mathrm{aw}}(a)) = \omega_0(a) \ .$$

⊣

Theorem 1.19 is a first instance of the correspondence principle for quantum fields, showing that regular quantum states have, in the limit $\hbar \to 0$, a classical counterpart, itself a regular state albeit with a possible loss of mass. Two questions arise naturally, in view of Theorem 1.19:

- Are cylindrical measures truly the physical classical states arising from the correspondence principle?

- Can we ensure that no mass is lost in the limit $\hbar \to 0$, at least for suitable families $\{\omega_\hbar \in \mathbb{W}_\hbar(\mathscr{T}, \varsigma)$ , $\hbar \in (0, \underline{\hbar})\}$ of quantum states?

The answer to the first question is that indeed cylindrical measures are the physically relevant set of classical states. First of all, cylindrical measures arise naturally in studying the correspondence principle for equilibrium states [see, *e.g.*, FKSS17, FKSS21, LNR15b, LNR18, LNR21]; furthermore, all cylindrical measures can be reached in the semiclassical limit, in the sense that given a cylindrical measure it is always possible to construct a family of quantum states that has such measure as a Wigner measure [see Fal18b, Thm. 1.11]. We prove this last result below in a simplified way, see Proposition 1.26.

Beforehand, let us discuss a different type of convergence to Wigner measures, that ensures that no mass is lost in the limit.

**Definition 1.21** (Convergence in the sense of Fourier transforms)**.**
Let $\{\omega_{\hbar_\beta} \in \mathrm{Reg}_{\hbar_\beta}(\mathscr{T}, \varsigma)_+$ , $\hbar_\beta \to 0\}$ be a net of quantum regular states, and let $\omega_0 \in \mathrm{Reg}(\mathscr{T}, \varsigma)_+$ be a classical regular state. We say that $\omega_{\hbar_\beta}$ *converges to* $\omega_0$ *in the sense of Fourier transforms* − denoted by $\omega_{\hbar_\beta} \xrightarrow[\hbar_\beta \to 0]{\mathscr{F}} \omega_0$ − if and only if $\hat{\omega}_{\hbar_\beta}$ converges pointwise to $\hat{\omega}_0$: for all $f \in \mathscr{T}$,

$$\lim_{\hbar_\beta \to 0} \hat{\omega}_{\hbar_\beta}(f) = \hat{\omega}_0(f) \ .$$

*Remark* 1.22. Definition 1.21 entails that the mass $M = \hat{\omega}_0$ of the cylindrical measure $\omega_0$ satisfies

$$M = \lim_{\hbar_\beta \to 0} M_\hbar =: \lim_{\hbar_\beta \to 0} \hat{\omega}_\hbar(0) \ .$$

In other words, a family of normalized regular quantum states converges in the sense of Fourier transforms to a normalized regular classical state: *no mass is lost in the limit procedure.*

The convergence in the sense of Fourier transforms yields the convergence in the sense of Wigner measures of Theorem 1.19, as shown in the next lemma.

**Lemma 1.23.** *Let $\{\omega_\hbar \in \mathbb{W}_\hbar(\mathscr{T}, \varsigma)$ , $\hbar \in (0, \underline{\hbar})\}$ be a family of regular quantum states, such that the net $(\hbar_\beta)_{\beta \in I} \subset (0, \underline{\hbar})$ (with $\hbar_\beta \to 0$). If $\omega_{\hbar_\beta} \xrightarrow[\hbar_\beta \to 0]{\mathscr{F}} \omega_0$ , then $\omega_0 \in \mathrm{Wig}\{\omega_\hbar$ , $\hbar \in (0, \underline{\hbar})\}$ and for all $a \in \mathscr{F}^{-1} L^1(\mathscr{T}^\star)_{\mathrm{cyl}}$ ,*

$$\lim_{\hbar_\beta \to 0} \omega_{\hbar_\beta}(\mathsf{q}_{\hbar_\beta}^{\mathscr{T}}(a)) = \int_{\mathscr{T}^\star}^\bullet a(T) \mathrm{d}\omega_0(T) \ .$$

*Proof.* Since $\omega_{\hbar_\beta} \xrightarrow[\hbar_\beta \to 0]{\mathscr{F}} \omega_0$, we have that, for all $a \in \mathscr{F}^{-1} L^1(\mathscr{T}^\star)_{\mathrm{cyl}}$ with dual base $F$, by dominated convergence the following holds:

$$\lim_{\hbar_\beta \to 0} \omega_{\hbar_\beta}(\mathsf{q}_{\hbar_\beta}^{\mathscr{T}}(a)) = \lim_{\hbar_\beta \to 0} \int_F \hat{a}(k) \hat{\omega}_{\hbar_\beta}(k) \mathrm{d}k$$

$$= \int_F \hat{a}(k) \Big(\lim_{\hbar_\beta \to 0} \hat{\omega}_{\hbar_\beta}(k)\Big) \mathrm{d}k = \int_F \hat{a}(k) \hat{\omega}_0(k) \mathrm{d}k = \int_{\mathscr{T}^\star}^\bullet a(T) \mathrm{d}\omega_0(T) \ .$$

$\dashv$

Let us conclude this section by studying a special class of regular states, called *coherent states*, since they will play an important role also in the study of the van Hove model. Let us specify to the concrete Euclidean case $\mathfrak{h} = \mathscr{S}(\mathbb{R}^d)$, and $\mathfrak{h}^\star = \mathscr{S}'(\mathbb{R}^d)$ with bilinear pairing $\mathrm{Re}\langle \cdot, \cdot \rangle_2 : \mathscr{S} \times \mathscr{S}' \to \mathbb{R}$.

**Definition 1.24** (Coherent states)**.**
Let $\underline{T} \in \mathscr{S}'$, and define $C_h(\underline{T}) \in \mathrm{Reg}_h(\mathfrak{h}, \langle \cdot, \cdot \rangle)_{+,1}$ – the *coherent state centered around* $\underline{T}$ – by its noncommutative Fourier transform

$$(\widehat{C_h(\underline{T})})(f) := e^{-\frac{\pi^2 h}{2}\langle f, f \rangle_2}\, W_0(f)[\underline{T}] = e^{-\frac{\pi^2 h}{2}\langle f, f \rangle_2}\, e^{2\pi i \mathrm{Re}\langle f, T \rangle_2} \ .$$

The coherent state centered around zero, $C_h(0)$, whose noncommutative Fourier transform is a Gaussian, is called the *Fock vacuum*, and it is the physical ground state of relativistic free field theories [see, *e.g.*, BSZ92, DG13, GJ87, Haa92, SW00].

It is not difficult to check, and we leave it to the reader, that $(\widehat{C_h(\underline{T})}) \in \mathfrak{G}_h(1)$.

**Lemma 1.25.** *Let us denote by $\mathscr{P}(\mathscr{S}')$ the set of Radon probability measures on $(\mathscr{S}', \beta(\mathscr{S}', \mathscr{S}))$, and let $\underline{T} \in \mathscr{S}'$. Furthermore, let $\delta_{\underline{T}} \in \mathscr{P}(\mathscr{S}')$ be the Dirac measures concentrated in $\underline{T}$. Then,*

$$C_h(\underline{T}) \xrightarrow[h \to 0]{\mathscr{F}} \delta_{\underline{T}} \ .$$

*Proof.* The proof is straightforward:

$$(\widehat{C_h(\underline{T})})(f) = e^{-\frac{\pi^2 h}{2}\langle f, f \rangle_2}\, W_0(f)[\underline{T}] \xrightarrow[h \to 0]{} W_0(f)[\underline{T}] = \hat{\delta}_{\underline{T}}(f) \ .$$

$\dashv$

The coherent states are quantum states of minimal uncertainty (provable in concrete representations), that converge to pure classical states, *i.e.* measures concentrated in a single phase-space point. They can be seen as a minimal $h$-deformation of $\delta_{\underline{T}}$ in the same spirit as the abstract quantization: from the convolution formula $\delta_0 * \delta_{\underline{T}} = \delta_{\underline{T}}$, we just deform $\delta_0$ to the noncommutative Fock vacuum (and recall that convolution maps to a product under the Fourier transform also for cylindrical measures). We can take this minimal deformation a step further, to produce for any $\mathfrak{m} \in \mathscr{M}(\mathscr{T}^*)_{\mathrm{cyl}}$ a minimally deformed quantum counterpart (for the sake of simplicity, we state the result for Euclidean spaces, but fixing a symplectic basis it could be extended to any symplectic space).

**Proposition 1.26** ([Fal18b, Thm. 1.11])**.** *For any $\underline{h} > 0$, the set of all Wigner measures corresponding to regular states in $\{\mathrm{Reg}_h(\mathfrak{h}, \langle \cdot, \cdot \rangle)_+ \, , \ h \in (0, \underline{h}]\}$ is $\mathscr{M}(\mathfrak{h}^*)_{\mathrm{cyl}}$.*

*More precisely, given $\mathfrak{m} \in \mathscr{M}(\mathfrak{h}^*)_{\mathrm{cyl}}$, the state $\omega_h(\mathfrak{m}) \in \mathrm{Reg}_h(\mathfrak{h}, \langle \cdot, \cdot \rangle)_+$ defined by*

$$(\widehat{\omega_h(\mathfrak{m})})(f) := e^{-\frac{\pi^2 h}{2}|f|^2}\, \hat{\mathfrak{m}}(f)$$

*converges to $\mathfrak{m}$ in the sense of Fourier transforms:*

$$\omega_h(\mathfrak{m}) \xrightarrow[h \to 0]{\mathscr{F}} \mathfrak{m} \ .$$

Again, we leave to the reader to check that indeed $(\widehat{\omega_h(\mathfrak{m})}) \in \mathfrak{G}_h(\mathfrak{m}(\mathfrak{h}^*))$ (it follows from the fact that the product of Fourier transforms is positive definite, and thus the product of a noncommutative and a commutative Fourier transform is quantum positive definite).

**1.5. The CCR-algebra and Wigner measures in the Fock representation.** Let us conclude this section by particularizing the above construction to free (time-zero) quantum fields, suitable to study non-relativistic quantum field theories in which a field interacts with non-relativistic particles, and that will be the setting for the van Hove model, to be discussed in § 2 and 3. The free field representation has been widely studied in the literature [see, *e.g.*, BR97, Coo51, Seg56, and references therein].

Throughout this section, let us specialize $\mathscr{T} = \mathscr{S}(\mathbb{R}^d)$, to be seen as a complex inner product space under the inner product $\langle \cdot, \cdot \rangle_2$. As discussed above in § 1.2, a complex inner product space is canonically isomorphic to a real symplectic and inner product space by the following identifications: $(\mathscr{S}(\mathbb{R}^d)_{\mathbb{R}}, \mathrm{Im}\langle \cdot, \cdot \rangle_2) \in$

**Symp**$_\mathbb{R}$, where $\mathscr{S}(\mathbb{R}^d)_\mathbb{R} \in \mathbf{Vec}_\mathbb{R}$ is $\mathscr{S}(\mathbb{R}^d)$ under the application of the forgetful functor that "forgets" about the complex structure (given $f \in \mathscr{S}(\mathbb{R}^d)$, $f$ and $if$ are real linearly independent in $\mathscr{S}(\mathbb{R}^d)_\mathbb{R}$); while $(\mathscr{S}(\mathbb{R}^d)_\mathbb{R}, \mathrm{Re}\langle\,\cdot\,,\,\cdot\,\rangle_2) \in \mathbf{Inn}_\mathbb{R}$. In choosing $\mathscr{T} = \mathscr{S}(\mathbb{R}^d)$, we mean with a slight abuse of notation that we are setting $(\mathscr{T}, \varsigma) \equiv (\mathscr{S}(\mathbb{R}^d)_\mathbb{R}, \mathrm{Im}\langle\,\cdot\,,\,\cdot\,\rangle_2)$. Furthermore, let us choose $\mathscr{T}^\star = \mathscr{S}'(\mathbb{R}^d)$ to be the phase space of time-zero fields (with the same abuse of notation), with the natural real bilinear pairing $\mathrm{Re}\langle\,\cdot\,,\,\cdot\,\rangle_2 : \mathscr{S} \times \mathscr{S}' \to \mathbb{R}$.

With the above choices, the Fock vacuum $C_h(0) \in \mathrm{Reg}_h(\mathscr{S}, \langle\,\cdot\,,\,\cdot\,\rangle_2)_{+,1}$ (see Definition 1.24) is defined by

$$(\widehat{C_h(0)})(f) = e^{-\frac{\pi^2 h}{2}\langle f,f\rangle_2} \ .$$

The GNS representation of $\mathbb{W}_h(\mathscr{S}, \langle\,\cdot\,,\,\cdot\,\rangle_2)$ induced by $C_h(0)$ is irreducible, and it is called the Fock representation (it is the *unique* – up to *-isomorphisms – irreducible representation if we substitute $\mathscr{S}$ with the finite dimensional $\mathbb{C}^d$), and it is customarily defined as follows, through the so-called *symmetric second quantization functor* $\Gamma_s$. The second quantization functor $\Gamma_s : \mathbf{Hilb}_\mathbb{C} \to \mathbf{Hilb}_\mathbb{C}$ is defined by its action on Hilbert spaces:

$$\Gamma_s(\mathscr{H}) = \oplus_{n\in\mathbb{N}} \mathscr{H}_n \ ,$$

with $\mathscr{H}_0 = \mathbb{C}$, and for any $n \geq 1$,

$$\mathscr{H}_n = \underbrace{\mathscr{H} \otimes_s \cdots \otimes_s \mathscr{H}}_{n} \ ,$$

where $\otimes_s$ denotes the symmetric tensor product. Its action on morphisms (unitary operators) is as follows: let $U \in \mathscr{L}(\mathscr{H})$ be a unitary operator on $\mathscr{H}$, then $\Gamma(U)$ is defined[3] by its action on vectors $\Psi \in \Gamma_s(\mathscr{H})$, $\Psi = (\Psi_0, \ldots, \Psi_n, \ldots)$ with $\Psi_n \in \mathscr{H}_n$:

$$(\Gamma(U)\Psi)_n = (U \otimes \cdots \otimes U)\Psi_n \ .$$

Given a densely defined self-adjoint operator $S$ on $\mathscr{H}$, its second quantization $\mathrm{d}\Gamma(S)$ is a densely defined self-adjoint operator on $\Gamma_s(\mathscr{H})$ defined through the second quantization functor by "passing to generators":

$$\Gamma(e^{-iS}) = e^{-i\mathrm{d}\Gamma(S)} \ ,$$

yielding, on the natural dense domain of $\mathrm{d}\Gamma(S)$,

$$(\mathrm{d}\Gamma(S)\Psi)_n = \sum_{j=1}^{n} S\big|_j \Psi \ ,$$

where $S\big|_j = \mathrm{id} \otimes \cdots \otimes \underset{j}{S} \otimes \cdots \otimes \mathrm{id}$ is the action of $S$ on the $j$-th copy of $\mathscr{H}$ in $\mathscr{H}_n$. The physical interpretation of the Fock space $\Gamma_s(\mathscr{H})$ is that its vectors are the wavefunctions of a quantum field with an arbitrary number of excitations described by $\mathscr{H}$: the fiber $\Psi_n \in \mathscr{H}_n$ gives the probability amplitude of having exactly $n$ excitations of the field with configuration $\Psi_n$. The Fock space has norm

$$\|\Psi\|^2 = \sum_{n\in\mathbb{N}} \|\Psi_n\|^2_{\mathscr{H}_n} \ ,$$

and therefore it is well suited only to describe fields with an "essentially finite" number of excitations: the probability of having $n$ excitations must go to zero as $n \to \infty$, otherwise the Fock norm cannot be finite. As we will discuss in great detail in § 2, there are many interesting physical situations in which the quantum field has an *infinite number of excitations* even in its fundamental state. Such systems *cannot be described* in the Fock representation, that in fact is not the unique irreducible representation of $\mathbb{W}_h(\mathscr{T}, \varsigma)$ whenever $\mathscr{T}$ is infinite dimensional (and this underlines the importance of developing an abstract, representation-independent, framework for semiclassical analysis). The Fock vacuum state $C_h(0)$ is represented by the projection on the Fock vacuum vector $\Omega_h = (1, 0, \ldots, 0, \ldots)$ with no field's excitations (hence the name). To define the representation of $\mathbb{W}_h(\mathscr{S}, \operatorname{Im}\langle\,\cdot\,,\,\cdot\,\rangle_2)$ on $\Gamma_s(L^2(\mathbb{R}^d))$, we need to introduce the very well-known *creation and annihilation operators*; these operators connect the fibers $\mathscr{H}_n \to \mathscr{H}_{n\pm1}$, hence their name. Beforehand, let us observe that the second quantization of the identity $\mathrm{d}\Gamma(1)$ is called the *number operator*, for it counts the number of excitations in each fiber $((\mathrm{d}\Gamma(1)\Psi)_n = n\Psi_n)$. The creation and annihilation operators $a^*(f)$ and $a(f)$, $f \in L^2(\mathbb{R}^d)$, are closed and densely defined operators, one adjoint to the other, with $D(\mathrm{d}\Gamma(1)^{\frac{1}{2}})$ as common core for all $f$ (see the bounds below). Their definition is customarily given as follows: let $K_n = \{k_1, \ldots, k_n\}$

$$(a^*(f)\Psi)_n(K_n) = \frac{1}{\sqrt{n}}\sum_{j=1}^n f(k_j)\Psi_{n-1}(K_n \smallsetminus \{k_j\})$$

$$(a(f)\Psi)_n(K_n) = \sqrt{n+1}\int_{\mathbb{R}^d}\bar{f}(k)\Psi_{n+1}(k, K_n)\mathrm{d}k$$

---

[3]for the action on morphisms, it is customary to omit the subscript $_s$ in the second quantization functor.

The creation and annihilation operators satisfy *canonical commutation relations* on $D(\mathrm{d}\Gamma(1))$:

$$[a(f), a^*(g)] = \langle f, g\rangle_2 \;.$$

The semiclassical parameter is introduced on the creation and annihilation, second quantized and $\Gamma$ operators as follows. Let us define the semiclassical creation and annihilation operators as

$$a_h^*(f) = a^*(\sqrt{h}f) \;,\; a_h(f) = a(\sqrt{h}f) \;;$$

the second quantized operators thus become[4]

$$\mathrm{d}\Gamma_h(S) = \mathrm{d}\Gamma(hS) \;.$$

The $\Gamma$ functor is typically kept $h$ independent, but for convenience of notation we define, for any strongly continuous unitary group $\{e^{-itS} \;,\; t \in \mathbb{R}\}$ in $L^2(\mathbb{R}^d)$,

$$\Gamma_h(e^{-itS}) = e^{-i\frac{t}{h}\mathrm{d}\Gamma_h(S)} \;.$$

The self-adjoint *Segal field* is defined, for all $f \in L^2(\mathbb{R}^d)$, as

$$(5) \qquad\qquad \varphi_h(f) = a_h^*(f) + a_h(f) \;.$$

Its unitary exponential

$$(6) \qquad\qquad W_h(f) := e^{i\pi\varphi_h(f)}$$

is called the *Weyl operator* and the slight abuse of notation with the noncommutative characters is justified as follows.

**Proposition 1.27.** *The* Fock *representation is the triple* $\left(\Gamma_s(L^2(\mathbb{R}^d)), \pi_F, \Omega_h\right)$, *where* $\pi_F$ *is defined by* $\pi_F(W_h(f)) = e^{i\pi\varphi_h(f)}$ *for any* $f \in \mathscr{S}$, *and it is an irreducible representation of* $\mathbb{W}_h(\mathscr{S}, \langle\,\cdot\,,\,\cdot\,\rangle_2)$ *with* $\Omega_h$ *as a cyclic vector.*

*Since*

$$\langle\Omega_h, e^{i\pi\varphi_h(f)}\Omega_h\rangle = (\widehat{C_h(0)})(f) \;,$$

*the Fock representation is the GNS-representation of the Fock vacuum* $C_h(0)$, *that is thus a* pure *state on* $\mathbb{W}_h(\mathscr{S}, \langle\,\cdot\,,\,\cdot\,\rangle_2)$. *For any* $g \in L^2(\mathbb{R}^d) \subset \mathscr{S}'(\mathbb{R}^d)$, *the coherent states* $C_h(g)$ *are Fock-normal and* pure: *the vector in* $\Gamma_s(L^2(\mathbb{R}^d))$ *representing them is called the coherent vector*

$$|C_h(g)\rangle = W_h(g/\pi ih)\Omega_h \;.$$

In the following lemma, whose proof is well-established, we collect some properties of creation and annihilation operators that will be useful in § 2.2.

---

[4] It is not difficult to show that if $\{e_j\}_{j\in\mathbb{N}} \subset D(S)$ is an orthonormal basis of $L^2(\mathbb{R}^d)$, then $\mathrm{d}\Gamma(S) = \sum_{j,k\in\mathbb{N}}\langle e_j, Se_k\rangle_2 a^*(e_j)a(e_k)$.

**Lemma 1.28.** *Let $f \in D(S)$, $S \geq 0$ on $\mathscr{H}$ with (possibly unbounded but densely defined and positive) inverse. Then,*

$$W_h(f/\pi i)^* d\Gamma_h(S) W_h(f/\pi i) = d\Gamma_h(S) + \hbar a_h^*(Sf) + \hbar a_h(Sf) + \hbar^2 \langle f, Sf \rangle_{\mathscr{H}} \,.$$

*Furthermore, for any $g$ such that $S^{-\frac{1}{2}} g \in \mathscr{H}$ and any $\Psi \in D(d\Gamma_h(S)^{\frac{1}{2}})$:*

$$\|a_h(g)\Psi\| \leq \|S^{-\frac{1}{2}} g\|_{\mathscr{H}} \, \|d\Gamma_h(S)^{\frac{1}{2}} \Psi\| \,.$$

*If, in addition, $g \in \mathscr{H}$ as well, then*

$$\|a_h^*(g)\Psi\| \leq \|S^{-\frac{1}{2}} g\|_{\mathscr{H}} \, \|d\Gamma_h(S)^{\frac{1}{2}} \Psi\| + \hbar \|g\|_{\mathscr{H}} \|\Psi\| \,.$$

For Fock-normal (thus also regular) states $\omega_{\varrho_h}$, stemming from a density matrix $\varrho_h \in \mathscr{L}^1\big(\Gamma_s(L^2(\mathbb{R}^d))\big)_{+,1}$ – $\mathscr{L}^1(\,\cdot\,)_{+,1}$ being the set of positive and norm one operators on $(\,\cdot\,)$ in the Schatten trace class ideal – there is a natural *sufficient condition* for convergence in the sense of Fourier transforms. Such condition could be reformulated more abstractly to hold also for suitable non-Fock states, but in view of our purposes and for ease of exposition let us limit to the Fock-normal case.

**Proposition 1.29** ([AN08]). *Let $\big\{\varrho_h \in \mathscr{L}^1\big(\Gamma_s(L^2(\mathbb{R}^d))\big)_{+,1}$, $\hbar \in (0, \underline{h})\big\}$ be a family such that $\exists \delta > 0$, $\exists C > 0$ such that for all $\hbar \in (0, \underline{h})$,*

$$\tag{7} \mathrm{Tr}\big(\varrho_h d\Gamma_h(1)^\delta\big) \leq C \,.$$

*Then there exists a sequence $\hbar_n \to 0$ and a Borel Radon measure $\mu \in \mathscr{P}(L^2(\mathbb{R}^d))$ such that $\varrho_h \xrightarrow[\hbar_n \to 0]{\mathscr{F}} \mu$. Here, the non-commutative Fourier transform takes the form: $\hat{\varrho}_h(f) = \mathrm{Tr}\big(\varrho_h W_h(f)\big)$.*

*Remark* 1.30. A few comments are in order:

- In view of (9)-(10), the representation $\big(\Gamma_s(L^2(\mathbb{R}^d)), \pi_F, \Omega_h\big)$ can be easily extended to a cyclic irreducible representation of $\mathbb{W}_h(L^2, \langle\,\cdot\,,\,\cdot\,\rangle_2)$, in which $\mathbb{W}_h(\mathscr{S}, \langle\,\cdot\,,\,\cdot\,\rangle_2)$ is a subalgebra. The density matrices $\varrho_h$ that satisfy (8) then clearly yield (regular) states on both algebras.

- When viewed as states of both algebras, they have a uniformly (with respect to $\hbar$) continuous Fourier transform (in both $\mathscr{S}$ and $L^2$ topologies), and by a diagonal extraction argument there exists the sequence $\hbar_n \to 0$ for the convergence in the sense of Fourier transforms. The resulting Fourier transform is also continuous in both topologies. By Bochner-Minlos theorem [see, *e.g.*, VTC87, Thm. 4.2], the corresponding cylindrical measure on $\mathscr{S}'$ is tight, and thus a $\beta(\mathscr{S}', \mathscr{S})$-Radon measure on $\mathscr{S}'$.

- When seen as a cylindrical measure on $L^2$, continuity of the Fourier transform does not imply tightness by itself. Ammari and Nier [AN08] however prove that any cylindrical measure that is a limit point for the convergence in the sense of Fourier transforms of a state satisfying (8) is indeed Prokhorov-tight and thus a Borel Radon measure on $L^2$.

- The two results clearly agree, since the measure $\mu$ on $L^2$ can bee seen as a Radon measure $\mu$ on $\mathscr{S}'$ that *concentrates on* $L^2 \subset \mathscr{S}'$: $\mu(\mathscr{S}' \smallsetminus L^2) = 0$.

## 2. The van Hove model, classical and quantum

In this section we introduce the *van Hove model* – or *van Hove-Miyatake model* [Miy52, Van52] – a toy model describing a scalar (relativistic) field that is created and annihilated by an immovable source of charge (*e.g.*, a very heavy ion). This toy model is – as discussed below – completely solvable, nonetheless it presents many features of realistic quantum field theories, especially concerning *infrared divergences*. The mathematical study of infrared and more general spectral problems in (semi- and non-relativistic) QFT stems from a seminal paper by Fröhlich [Frö73], and has developed across this and the last century into a prolific and active field of research, with several new key ideas with application far beyond its original scope [it would be too cumbersome to cite all results here, let me limit to an inconclusive selection: AH12, Amm00, Amm04, AF13, Ara01, Ara20, AH97, AHH99, BBIP21, BBP17, Bac+13, BFP06, BFS98a, BFS98b, BFS99, BFSS99, BH22, BDP12, BFFS15, BCV03, BDG04, Bet+02, BZ21, CFP10, DM20a, DM20b, DP12, DG99, DJP03, DJ03, FS14, FMS04, FKS09, FM04a, FM04b, Gér02, GHPS12, GP09, GHPS11, GLL01, Gro72, HHS24, Hir06, HH14, HS01a, HS01b, Hir99, Hir01, HM22, HS13, HS05, JP02, LMM23, MSB07, Møl05, Møl06, Par04, Par08, Par10, Piz03, Pol23, RW23, Sch06, Spo89, Wak16, see also the references therein for additional important contributions]. In focusing on the van Hove model, we follow in many aspects closely [Ara20, Der03], as well as [Frö73] for the C*-algebraic point of view.

The rest of this section is organized as follows: in § 2.1 we introduce the classical version of the van Hove model, as an infinite dimensional mechanical system of Hamiltonian type; in § 2.2 we focus on its "standard" quantized version in the Fock representation, and on the so-called *infrared catastrophe*; finally, in § 2.3 we reformulate both the classical and quantum van Hove models as algebraic dynamical map, an approach better suited at the study of the correspondence principle $\hbar \to 0$ by abstract semiclassical analysis, that is performed in § 3 below.

## 2.1. The classical van Hove model.
In this section we study a classical scalar field that is created and absorbed by a static charge distribution. Let us denote by $\mathscr{S}_0$ the subset of Schwartz functions supported on $\mathbb{R}^d \setminus \{0\}$, and by $\alpha \in \mathscr{S}'_0(\mathbb{R}^d)$ the scalar field in momentum representation, by $\varpi(k) = \sqrt{k^2 + \mu^2}$, $\mu \geq 0$, the Klein-Gordon dispersion relation with mass $\mu \geq 0$, and by $j \in \mathscr{S}'_0(\mathbb{R}^d)$ the charge distribution in momentum space[5]. The *classical van Hove equation* is then the linear inhomogeneous equation on $\mathscr{S}'_0$:

$$i\partial_t \alpha - \varpi \alpha = j \,. \tag{8}$$

Introducing the subset $\mathscr{S}'_{1/\varpi}(\mathbb{R}^d) := \{\alpha \in \mathscr{S}'_0(\mathbb{R}^d) \,, \ \varpi^{-\frac{1}{2}} \alpha \in \mathscr{S}'(\mathbb{R}^d)\}$, we can define the scalar field in position representation $A \in \mathscr{S}'(\mathbb{R}^d)$ as

$$A = \sqrt{2}\mathrm{Re}\mathscr{F}^{-1}(\varpi^{-\frac{1}{2}} \alpha) \,. \tag{9}$$

The field in position representation satisfies the well-known *inhomogoeneous Klein-Gordon equation*

$$(\Box + \mu^2)A = j \,, \tag{10}$$

where $\Box = \partial_t^2 - \Delta$, and

$$j = 2\sqrt{2}\mathrm{Im}\mathscr{F}^{-1}(\varpi^{-\frac{1}{2}} j) \,,$$

for any source $j$ of the form above with $j \in \mathscr{S}'_{1/\varpi}$. In all our concrete cases of interest, (9) is invertible, and therefore (10) and (8) are equivalent, in the sense that one is well-posed in some suitable space if and only if the other is well-posed in the corresponding space, as given by (9) itself. As it will become apparent in § 2.3 and § 3, while (10) is the standard version one can find in textbooks, (8) is the most natural to study the correspondence principle.

Now, we would like to see (8) as the Hamilton equation yielded by an energy functional $\mathscr{E}$, that we will call the *classical van Hove energy*. In order to do that, let us define the energy space for the van Hove equation to be

$$L^2_\varpi(\mathbb{R}^d) := \{\alpha \in \mathscr{S}'_0(\mathbb{R}^d) \,, \ \sqrt{\varpi} \alpha \in L^2(\mathbb{R}^d)\} \,. \tag{11}$$

---

[5]The dispersion relation $\varpi$ can be effortlessly generalized to a different function or even a suitable linear operator; we stick to the case of a Klein-Gordon field to improve readability.

Let us also define the source spaces

$$
\begin{aligned}
L^2_{1/\varpi}(\mathbb{R}^d) &:= \{ J \in \mathscr{S}'_0(\mathbb{R}^d) \ , \ \varpi^{-\frac{1}{2}} J \in L^2(\mathbb{R}^d) \} \ ; \\
L^2_{1/\varpi^2}(\mathbb{R}^d) &:= \{ J \in \mathscr{S}'_0(\mathbb{R}^d) \ , \ \varpi^{-1} J \in L^2(\mathbb{R}^d) \} \ .
\end{aligned}
$$

(12)

**Definition 2.1** (The classical van Hove energy)**.**

For any source $J \in L^2_{1/\varpi}$, the *classical van Hove energy* $\mathscr{E} : L^2_\varpi \to \mathbb{R}$ is defined as

$$
\mathscr{E}(\alpha) := \langle \alpha, \varpi \alpha \rangle_2 + \langle \alpha, J \rangle_2 + \langle J, \alpha \rangle_2 = \|\alpha\|^2_{L^2_\varpi} + 2\mathrm{Re}\langle \alpha, J \rangle_2 \ .
$$

The Hamilton equation of $\mathscr{E}$ is

$$
i\partial_t \alpha = \varpi \alpha + J \ ,
$$

equivalent to the inhomogeneous Klein-Gordon equation

$$
(\Box + \mu^2) A = j \ .
$$

*Remark* 2.2. $\mathscr{E}$ is defined on the whole $L^2_\varpi$ since, for any $J \in L^2_{1/\varpi}$, it can be indeed rewritten as

$$
\mathscr{E}(\alpha) = \|\alpha\|^2_{L^2_\varpi} + 2\mathrm{Re}\langle \sqrt{\varpi} \alpha, J / \sqrt{\varpi} \rangle_2 \ .
$$

If $J \in L^2 \smallsetminus L^2_{1/\varpi}$, $\mathscr{E}$ can still be densely defined in $L^2_\varpi$ (it is defined on $L^2 \cap L^2_\varpi$), however in that case it can be shown to be *unbounded from below* (see § 2.2 below for additional comments on such "infrared singularities").

For any $J \in L^2_{1/\varpi}$, the Cauchy problem for the van Hove equation (with initial datum $\alpha_0$) has an explicit unique solution on $L^2_\varpi$, namely

$$
\alpha(t) = e^{-it\varpi} \left( \alpha_0 + \frac{J}{\varpi} \right) - \frac{J}{\varpi} \ ,
$$

(13)

observing that for any $J \in L^2_{1/\varpi}, J/\varpi \in L^2_\varpi$.

**Definition 2.3** (The van Hove Hamiltonian flow)**.**

For any source $J \in L^2_{1/\varpi}$, the *van Hove Hamiltonian flow* $\Phi_t : L^2_\varpi \to L^2_\varpi$ is defined by

$$
\Phi_t(\alpha) = e^{-it\varpi} \left( \alpha_0 + \frac{J}{\varpi} \right) - \frac{J}{\varpi} \ .
$$

The Hamiltonian $\mathscr{E}$ is a constant of motion for $\Phi_t$: for any $t \in \mathbb{R}$,

$$
\mathscr{E} \circ \Phi_t = \mathscr{E} \ .
$$

The van Hove energy is "explicitly diagonalizable", so its boundedness from below and minimizer is obvious.

**Lemma 2.4.** *For any $J \in L^2_{1/\varpi}$,*

$$
\inf_{\alpha \in L^2_\varpi} \mathscr{E}(\alpha) = -\|J\|^2_{L^2_{1/\varpi}} \ ;
$$

*furthermore, $\alpha_{\mathrm{gs}} = -J/\varpi$ is the unique minimizer of $\mathscr{E}$: $-J/\varpi \in L^2_\varpi$, and*

$$
\mathscr{E}(-J/\varpi) = -\|J\|^2_{L^2_{1/\varpi}} \ .
$$

*Finally, the minimizer is also stationary: for any $t \in \mathbb{R}$, $\Phi_t(-J/\varpi) = -J/\varpi$.*

*Proof.* By completing the square, $\mathscr{E}$ can be rewritten as

$$
\mathscr{E}(\alpha) = \|\alpha + J/\varpi\|^2_{L^2_\varpi} - \|J\|^2_{L^2_{1/\varpi}} \ .
$$

(14)

Since $L^2_\varpi$ is a Hilbert space, $\|\beta\|_{L^2_\varpi} = 0$ if and only if $\beta = 0$, that yields both the existence and uniqueness of a minimizer and the ground state energy. The fact that the minimizer of an energy is stationary with respect to the corresponding Hamiltonian flow shall come at no surprise, however it can also be checked explicitly and straightforwardly using Definition 2.3. $\dashv$

The rewriting (14) leads also to the following observation on the Hamiltonian flow $\Phi_t$. Let $\Phi_t^0$ be the unitary flow on $L_\varpi^2$ defined by

$$\Phi_t^0(\alpha) = e^{-it\varpi} \alpha \ ,$$

and let $\Theta_\beta : L_\varpi^2 \to L_\varpi^2$ be the translation by some $\beta \in L_\varpi^2$:

$$\Theta_\beta(\alpha) = \alpha + \beta \ .$$

Such definitions yield trivially, in light of Definition 2.3, the following result.

**Lemma 2.5.** *For any $J \in L_{1/\varpi}^2$, the van Hove Hamiltonian flow can be rewritten as*

$$\Phi_t = \Theta_{-J/\varpi} \circ \Phi_t^0 \circ \Theta_{J/\varpi} \ .$$

This rewriting will be of particular usefulness in § 2.3 and § 3.

## 2.2. The quantum van Hove model in Fock representation.

In this section we study the quantum van Hove model in the "orthodox" way, by resorting to its concrete definition in the Fock representation. In the Fock representation, the van Hove Hamiltonian can be seen as the formal normal ordered quantization of the classical van Hove energy $\mathscr{E}$ (see Definition 2.1), in which each function $\alpha$ is replaced by the annihilation operator $a$, and each $\bar{\alpha}$ by $a^*$, putting all creations to the left of annihilations. The (formal) quantum van Hove Hamiltonian thus reads

$$(15) \qquad\qquad H_h = \mathrm{d}\Gamma_h(\varpi) + a_h^*(J) + a_h(J) \ .$$

The main interest in the quantum van Hove model stems from its infrared behavior, in dependence of the infrared regularity of the source $J$. Contrarily to the classical case, there is no unified treatment of the existence of minimizers for the quantum van Hove model. Before discussing such infrared behavior, let us observe that for the *massive field* $\mu > 0$ there can be no infrared singularity apart from eventual unboundedness from below due to the mass gap; the physically interesting case is that of a *massless field* $\mu = 0$, yielding a dispersion relation $\varpi(k) = |k|$. While the discussion below is valid for a general value of the mass (and even for more general choices of $\varpi$), the reader shall keep in mind the massless field as a reference. The infrared behavior of the quantum van Hove model in Fock representation is given by the following theorem [see Ara20, Der03]. Beforehand, let us denote by $\mathscr{C}_c^\infty(\mathrm{d}\Gamma_h(1)) = \{\Psi \in \Gamma_s(L^2(\mathbb{R}^d)) \ , \ \exists N \in \mathbb{N} \ , \ \Psi_n = 0 \ \forall n \geq N\}$ the set of *finite particle vectors*.

**Theorem 2.6.** *Let $H_h$ be the van Hove operator on $\Gamma_s(L^2(\mathbb{R}^d))$ formally defined by (15). The following three cases are possible:*

- *Infrared regular case:* $\boxed{J \in L_{1/\varpi}^2 \cap L_{1/\varpi^2}^2}$ . *$H_h$ is diagonalized by $W_h(J/\pi i\hbar\varpi)$:*

  $$H_h = W_h(J/\pi i\hbar\varpi)^* \Big( \mathrm{d}\Gamma_h(\varpi) - \|J\|_{L_{1/\varpi}^2}^2 \Big) W_h(J/\pi i\hbar\varpi) \ .$$

  *It thus follows that $H_h$ is self-adjoint on*

  $$W_h(J/\pi i\varpi) D(\mathrm{d}\Gamma_h(\varpi)) \subset D(\mathrm{d}\Gamma_h(\varpi)^{\frac{1}{2}}) \ ,$$

  *that*

  $$\inf \sigma(H_h) = -\|J\|_{L_{1/\varpi}^2} \ ,$$

  *and*

  $$|C_h(-J/\varpi)\rangle\langle C_h(-J/\varpi)| \text{ is its unique ground state.}$$

- Infrared singularity of type I: $\boxed{J \in L^2_{1/\varpi} \smallsetminus L^2_{1/\varpi^2}}$. $H_h$ is self-adjoint on $D(H_h) \subset D(\mathrm{d}\Gamma_h(\varpi)^{\frac{1}{2}})$,

$$\inf \sigma(H_h) = -\|J\|^2_{L^2_{1/\varpi}} \, ,$$

  and $H_h$ has no Fock-ground state.

- Infrared singularity of type II: $\boxed{J \in L^2 \smallsetminus L^2_{1/\varpi}}$. $H_h$ is essentially self-adjoint on $D(\mathrm{d}\Gamma_h(\varpi)) \cap \mathscr{C}^\infty_c(\mathrm{d}\Gamma_h(1))$, and

$$\inf \sigma(H_h) = -\infty \, .$$

*Proof.*

*Infrared regular case.* Since $J \in L^2_{1/\varpi^2}$, $W_h(J/\pi i\hbar\varpi)$ is well defined as a unitary operator on $\Gamma_s(L^2(\mathbb{R}^d))$. The translation properties of the coherent operators, see Lemma 2.16, yield

$$H_h = W_h(J/\pi i\hbar\varpi)^* \Big(\mathrm{d}\Gamma_h(\varpi) - \|J\|^2_{L^2_{1/\varpi}}\Big) W_h(J/\pi i\hbar\varpi) \, .$$

The spectrum of $\mathrm{d}\Gamma_h(\varpi)$ is $\sigma(\mathrm{d}\Gamma_h(\varpi)) = \mathbb{R}_+$, with zero as the only embedded eigenvalue, with the Fock vacuum $\Omega_h$ as eigenvector. Therefore, $\inf \sigma(H_h) = -\|J\|^2_{L^2_{1/\varpi}}$, and the coherent state $|C_h(-J/\varpi)\rangle\langle C_h(-J/\varpi)|$ is the unique ground state of $H_h$. To prove that the domain of $H_h$ is included in $D(\mathrm{d}\Gamma_h(\varpi)^{\frac{1}{2}})$, observe that thanks to Lemma 2.16 and $J \in L^2_{1/\varpi}$, the quadratic form of $H_h$ satisfies: for all $\Psi \in D(\mathrm{d}\Gamma_h(\varpi)^{\frac{1}{2}})$,

$$\Big|\big\langle \Psi, (a^*_h(J) + a_h(J))\Psi\big\rangle\Big| \le 2\|J\|_{L^2_{1/\varpi}} \big\|\mathrm{d}\Gamma_h(\varpi)^{\frac{1}{2}}\Psi\big\| \|\Psi\| \, .$$

By KLMN theorem [RS75, Theorem X.17] it follows that $D(H_h) \subset D(\mathrm{d}\Gamma_h(\varpi)^{\frac{1}{2}})$. Let us remark that if $f \in L^2_{1/\varpi} \cap L^2$, by Kato-Rellich theorem [RS75, Theorem X.12], $D(H_h) = D(\mathrm{d}\Gamma_h(\varpi))$.

*Infrared singularity of type I.* Self-adjointness follows from KLMN theorem, as discussed above, as well as boundedness from below with $\inf \sigma(H_h) \ge -\|J\|^2_{L^2_{1/\varpi}}$. On the other hand, by letting $\{J_n\}_{n\in\mathbb{N}} \subset L^2_{1/\varpi^2}$ be the infrared-regularized source defined by $J_n(k) = 1_{|\cdot| \ge \frac{1}{n}}(k) J(k)$, and choosing $\Psi^{(n)} = |C_h(-J_n/\varpi)\rangle$, one obtains

$$\langle \Psi^{(n)}, H_h \Psi^{(n)}\rangle = \|J_n\|^2_{L^2_{1/\varpi}} - 2\|J_n\|^2_{L^2_{1/\varpi}} = -\|J_n\|^2_{L^2_{1/\varpi}} \, .$$

Therefore, it follows that

$$\inf \sigma(H_h) \le \liminf_{n\to\infty}\langle \Psi^{(n)}, H_h \Psi^{(n)}\rangle = -\|J\|^2_{L^2_{1/\varpi}} \, .$$

It remains to prove that $H_h$ has no ground state. To do so, remark that we can split $L^2(\mathbb{R}^d) = L^2_\ge(n) \oplus L^2_\le(n)$, with

$$L^2_\ge(n) = 1_{|\cdot| \ge \frac{1}{n}}\left(L^2(\mathbb{R}^d)\right), \ L^2_\le(n) = (1 - 1_{|\cdot| \ge \frac{1}{n}})\left(L^2(\mathbb{R}^d)\right) \, .$$

Therefore, $\Gamma_s(L^2(\mathbb{R}^d)) \cong \Gamma_s(L^2_\ge(n)) \otimes \Gamma_s(L^2_\le(n))$, and

$$H_h \cong H_\le + H_\le = \left(\mathrm{d}\Gamma_h(\varpi) + a^*_h(J^\perp_n) + a_h(J^\perp_n)\right) \otimes 1 + 1 \otimes \left(\mathrm{d}\Gamma_h(\varpi) + a^*_h(J_n) + a_h(J_n)\right) \, ,$$

with $f^\perp_n = \left(1 - 1_{|\cdot| \ge \frac{1}{n}}\right)f$. Now suppose that $H_h$ has a ground state $\Psi$. Then $\Psi$ must be a ground state for both $H_\le$ and $H_\ge$. But since $H_\ge$ is a van Hove Hamiltonian with source $J_n \in L^2_{-1}$, its ground state vector must be of the form $\Psi = \Phi_\le \otimes |C_h(-J_n/\varpi)\rangle$. Take now a vector $g \in \bigcup_{n\ge 1} L^2_\le(n)$, densely embedded in $L^2(\mathbb{R}^d)$. By choosing the above decomposition for $n$ large enough — $n \ge \tilde{n}$, $\tilde{n}$ such that $g \in L^2_\le(\tilde{n})$ — we have that

$$a_h(g)\Psi = -\langle g, J/\varpi\rangle_2 \Psi \, .$$

Now, since $\bigcup_{n\ge 1} L^2_\le(n)$ is dense in $L^2$, this can only be true if either $J/\varpi \in L^2$ or $\Psi = 0$. This is absurd since by hypothesis $J/\varpi \notin L^2$, and $\langle \Psi, H_h\Psi\rangle\big|_{\Psi=0} = 0 > -\|J\|^2_{L^2_{1/\varpi}}$ for $J \ne 0$; therefore, $H_h$ has no ground state.

*Infrared singularity of type II.* Essential self-adjointness of $H_\hbar$ follows from Nelson's commutator theorem [RS75, Theorem X.36]. To prove that $H_\hbar$ is unbounded from below, let $\{J_n\}_{n\in\mathbb{N}} \subset L^2_{1/\varpi^2}$ be defined as above by $J_n(k) = 1_{|\cdot|\geq\frac{1}{n}}(k) J(k)$, and choose $\Psi^{(n)} = |C_\hbar(-J_n/\varpi)\rangle$. Here one obtains

$$\langle \Psi^{(n)}, H_\hbar \Psi^{(n)}\rangle = -\|J\|^2_{L^2_{1/\varpi}} \xrightarrow[n\to\infty]{} -\infty$$

since $J \notin L^2_{1/\varpi}$. ⊣

*Remark* 2.7. In light of Theorem 2.6 we can have some hints of how the correspondence principle works in the van Hove model:

- The quantum ground state energy $E_\hbar := \inf \sigma(H_\hbar) = -\|J\|^2_{L^2_{1/\varpi}}$ does not depend on $\hbar$, and it coincides with the ground state energy $E_0$ of the classical van Hove model, see Lemma 2.4. Therefore, the correspondence principle for ground state energies $\lim_{\hbar\to 0} E_\hbar = E_0$ is trivially satisfied in the van Hove model [in other more realistic models, such correspondence is not obvious and requires some effort to be proved, see AF14, CFO23a].

- The infrared properties of both the quantum and classical van Hove models depend on the regularity of the source $J$. However, while in both the infrared regular and II-infrared singular cases the quantum and classical models share the same features (they are both either bounded or unbounded from below, and in the former case they both have a unique ground state), in the I-infrared singular case they behave differently: the classical model has a unique ground state, while the quantum model has no (Fock-)ground state. As a matter of fact, this difference reflects a well-known purely quantum problem related to field theories: the emergence of *physical singularities* by quantization. In this case, the singularity is due to the fact that the ground state in the I-infrared singular case has infinitely many field excitations, as discussed below, making the Fock representation unsuited to describe it. As we prove in § 3 our representation-independent abstract semiclassical analysis is robust enough to be "transparent" to such a problem, making the correspondence principle hold in full generality for the van Hove model [this robustness of the correspondence principle seems rather general, in AF17, it is showed to hold also in the more difficult Nelson model with self-energy renormalization].

As discussed above, the infrared behavior of $H_\hbar$ in its region of self-adjointness $J \in L^2_{1/\varpi} \cup L^2$, as dictated by Theorem 2.6, paints a clear picture, namely that there are two regularity thresholds, one purely quantum (Fock), and one both classical and quantum: as soon as $J \notin L^2_{1/\varpi^2}$, the operator *loses its ground state*; as soon as $J \notin L^2_{1/\varpi}$, it becomes *unstable* (unbounded from below). The loss of a ground state takes the name of *infrared catastrophe* in the literature [see Ara20], albeit being itself less dramatic than the loss of stability. To better understand the infrared catastrophe from a physical perspective, it is instructive to track the dependence on the source on the van Hove Hamiltonian $H_\hbar(J)$, and study the behavior as $n \to \infty$ of the infrared regularized $H_\hbar(J_n)$ – with $J_n = 1_{|\cdot|\geq\frac{1}{n}} J$ – in presence of an infrared singularity of type I.

**Lemma 2.8.** *Let $J \in L^2_{1/\varpi} \setminus L^2_{1/\varpi^2}$, and let $J_n = 1_{|\cdot|\geq\frac{1}{n}} J$ be its infrared regularization. Furthermore, let $|C_\hbar(-J_n/\varpi)\rangle \in \Gamma_s(L^2(\mathbb{R}^d))$ be the ground state vector of $H_\hbar(J_n)$. Then:*

- $\lim_{n\to\infty}\langle C_\hbar(-J_n/\varpi), d\Gamma_\hbar(1) C_\hbar(-J_n/\varpi)\rangle = +\infty$

- $w\text{-}\lim_{n\to\infty}|C_\hbar(-J_n/\varpi)\rangle = 0$

*Proof.* The translation property of Weyl operators (see Lemma 2.16) yields

$$\langle C_\hbar(-J_n/\varpi), d\Gamma_\hbar(1) C_\hbar(-J_n/\varpi)\rangle = \|J_n\|^2_{L^2_{1/\varpi^2}} \xrightarrow[n\to\infty]{} +\infty \ .$$

Now, by the Weyl relations one gets, for $f \neq J_n/\pi i\hbar\varpi$,

$$\langle W_\hbar(f)\Omega_\hbar, W_\hbar(J_n/\pi i\hbar\varpi)^*\Omega_\hbar\rangle = e^{i\pi\mathrm{Re}\langle -J_n/\varpi, f\rangle_2}\langle W_\hbar(f + J_n/\pi i\hbar\varpi)\Omega_\hbar, \Omega_\hbar\rangle \ .$$

The expectation of Weyl operators on the Fock vacuum is given by Definition 1.24 and Proposition 1.27, yielding

$$\langle W_h(f)\Omega_h, W_h(J_n/\pi i\hbar\varpi)^*\Omega_h\rangle = e^{-\frac{\pi^2 h}{2}\|f+J_n/\pi i\hbar\varpi\|_2^2 + i\pi\mathrm{Re}\langle -J_n/\varpi, f\rangle_2} \ .$$

Hence,

$$\lim_{n\to\infty}\left|\langle W_h(f)\Omega_h, W_h(J_n/i\hbar\varpi)^*\Omega_h\rangle\right| = \lim_{n\to\infty} e^{-\frac{\pi^2 h}{2}\|f+J_n/i\hbar\varpi\|_2^2} = 0 \ .$$

Since $\Omega_h$ is cyclic for $\mathbb{W}_h(\mathscr{S}(\mathbb{R}^d))$, the set $\left\{W_h(f), f\in\mathscr{S}(\mathbb{R}^d)\ ,\ f\neq J_n/\pi i\hbar\varpi\right\}\Omega_h\subset\Gamma_s(L^2(\mathbb{R}^d))$ is total, that combined with the limit above proves that for all $\Gamma_s(L^2(\mathbb{R}^d))\ni\Psi\neq C_h(-J_n/\varpi)$, $\lim_{n\to\infty}\langle\Psi, C_h(-J_n/\varpi)\rangle = 0$. ⊣

The above Lemma 2.8 can be interpreted, physically, as follows. Suppose that there is a ground state $\omega_{h,\mathrm{gs}}$ for the van Hove model $H_h(J)$ with an infrared singularity of type I (as we discuss in § 2.3 below, a ground state indeed exists in a non-Fock representation). Such ground state is the weak-* limit $n\to\infty$ (in the algebraic sense), of the Fock-normal ground states $C_h(J_n/\varpi)$ of the regularized van Hove model $H_h(J_n)$ (again, see § 2.3 below for further details). By Lemma 2.8 it follows that the expectation of the Fock number operator on the Fock realization of $C_h(J_n/\varpi)$ *diverges* as the infrared cutoff is removed, and the transition amplitude between $C_h(J_n/\varpi)$ and any Fock state *vanishes*. This is interpreted as the fact that the ground state of the van Hove model has *an infinite number of Fock excitations*: since there is no mass gap $\mu = 0$, each excitation can have an infinitesimal energy, resulting in a finite total energy $(-\|J\|_{L^2_{1/\varpi}}^2)$. This infinite number of excitations with very low energy are called *soft photons*, and they are at the origin of the infrared catastrophe: as soon as the source $J\notin L^2_{1/\varpi^2}$, it becomes sufficiently singular at low momenta to create an infinite number of soft photons in the ground state, "kicking the latter out" of the Fock representation, that by construction can only accommodate states with zero probability of having infinitely many excitations (as discussed above in § 1.5).

## 2.3. The van Hove model as an algebraic dynamical system.

In this section we reformulate the van Hove model − classical and quantum − as an algebraic dynamical system, and study the properties of its dynamics, long time asymptotics and equilibrium states in the algebraic framework, paving the way for the study of the correspondence principle by abstract semiclassical analysis of § 3.

### 2.3.1. *The classical van Hove dynamical system.*

Let us define $\mathfrak{L} := L^2_{1/\varpi}\cap L^2$, and $\langle\cdot,\cdot\rangle_{\mathfrak{L}} = \langle\cdot,\cdot\rangle_2$, and observe that $(\mathfrak{L},\langle\cdot,\cdot\rangle_2)\in\mathbf{Inn}_\mathbb{C}$. Furthermore, let us choose $\mathfrak{L}^\star = L^2_\varpi$ with duality bracket $B(\cdot,\cdot) = \langle\cdot,\cdot\rangle_2$: this choice is possible since $\langle\cdot,\cdot\rangle_2 = \langle\varpi^{-1/2}\cdot,\varpi^{1/2}\cdot\rangle_2$ is separating points in $\mathfrak{L}$. Furthermore, $e^{it\varpi}$ is a unitary operator on $L^2_{\varpi^\alpha}$ for any $\alpha\in\mathbb{R}$, and thus leaves $\mathfrak{L}$ invariant. Hence, $e^{it\varpi}$ is a symplectic linear transformation on $(\mathfrak{L}_\mathbb{R}, \mathrm{Re}\langle\cdot,\cdot\rangle_{\mathfrak{L}})$.

Let us now consider the algebra of almost periodic functions $\mathbb{W}_0(\mathfrak{L},\langle\cdot,\cdot\rangle_{\mathfrak{L}})$. Given a character $W_0(f)$, let us define the van Hove dynamical map $\tau_0(t)$, in view of Lemma 2.5, as

$$\tau_0(t)\big[W_0(f)\big] = W_0(f)\circ\Phi_t = W_0(f)\circ\Theta_{-J/\varpi}\circ\Phi_t^0\circ\Theta_{J/\varpi} \ ,$$

for any source $J\in L^2_{1/\varpi}$. In other words, the character $W_0(f)$ gets mapped by $\tau_0(t)$ in the element of $\mathbb{W}_0(\mathfrak{L},\langle\cdot,\cdot\rangle_{\mathfrak{L}})$ given by:

$$\tau_0(t)\big[W_0(f)[T]\big] = W_0(f)\big[e^{-it\varpi}(T+J/\varpi)-J/\varpi\big] = W_0(e^{it\varpi}f)[T]e^{2\pi i\mathrm{Re}\langle f,(e^{-it\varpi}-1)J/\varpi\rangle_2} \ ,$$

recalling that $J/\varpi\in L^2_\varpi$, and thus $\langle f,(e^{-it\varpi}-1)J/\varpi\rangle_2$ is well-defined for any $f\in\mathfrak{L}\subset L^2_{1/\varpi}$. By linearity, the action extends to trigonometric polynomials as

$$\tau_0(t)\big[\Pi[T]\big] = \Pi\big[e^{-it\varpi}(T+J/\varpi)-J/\varpi\big] \ ,$$

clearly preserving the algebraic operations. Furthermore, since

$$\left\|\tau_0(t)[\Pi]\right\| = \sup_{T \in \mathfrak{L}} \left|\Pi\left[e^{-it\varpi}(T + J/\varpi) - J/\varpi\right]\right| = \left\|\Pi\right\| \,,$$

$\{\tau_0(t) \,,\ t \in \mathbb{R}\}$ extends by density to an isometric group of *-automorphisms of $\mathbb{W}_0(\mathfrak{L}, \langle\,\cdot\,,\,\cdot\,\rangle_{\mathfrak{L}})$.

**Definition 2.9** (Classical van Hove dynamical map)**.**
The isometric group of *-automorphisms $\{\tau_0(t) \,,\ t \in \mathbb{R}\}$ on $\mathbb{W}_0(\mathfrak{L}, \langle\,\cdot\,,\,\cdot\,\rangle_{\mathfrak{L}})$ is called the *classical van Hove dynamical map*.

Its transposed action $\tau_0(t)^{\mathrm{t}}$ on regular states $\omega_0 \in \mathrm{Reg}_0(\mathfrak{L}, \langle\,\cdot\,,\,\cdot\,\rangle_{\mathfrak{L}})_+$ is defined by

$$\left(\widehat{\tau_0(t)^{\mathrm{t}}[\omega_0]}\right)(f) = \hat{\omega}_0(e^{it\varpi}f)e^{2\pi i \mathrm{Re}\langle f, (e^{-it\varpi}-1)J/\varpi\rangle_2} = \left(\widehat{\Phi_{t\,*}\,\omega_0}\right)(f) \,,$$

where $\Phi_{t\,*}\,\omega_0$ denotes the *pushforward* of the regular state (cylindrical measure) by the van Hove flow $\Phi_t$.

When no confusion arises, we shall refer to the classical van Hove dynamical map only by $\tau_0$ (both when acting on observables and on states by transposition).

Given this dynamical map, one can study equilibrium states (in particular, the ground state) and long-time asymptotics from an algebraic viewpoint. Although it would probably be possible to be more general, let us restrict to two types of equilibrium states: the ground state, and the Gibbs states. We already saw that the ground state of the functional $\mathscr{E}$ is unique, and it is given by $-J/\varpi$.

**Lemma 2.10.** *For any source $J \in L^2_{1/\varpi}$, the normalized regular state $\delta_{-J/\varpi} \in \mathrm{Reg}_0(\mathfrak{L}, \langle\,\cdot\,,\,\cdot\,\rangle_{\mathfrak{L}})_{+,1}$ is an equilibrium state for $\tau_0$, and it is the unique $\beta(\mathscr{S}', \mathscr{S})$-Radon probability measure $\mu$ such that*

$$\int_{\mathscr{S}'} \mathscr{E}(T) \mathrm{d}\mu(T) = E_0 = -\|J\|^2_{L^2_{1/\varpi}} \,.$$

*We call $\delta_{-J/\varpi}$ the* ground state *of the classical van Hove dynamical map.*

*Proof.* The fact that $\delta_{-J/\varpi}$ is an equilibrium state follows immediately from an explicit computation using its Fourier transform $\hat{\delta}_{-J/\varpi}(f) = e^{2\pi i \mathrm{Re}\langle f, -J/\varpi\rangle_2}$.

Furthermore, since $-J/\varpi$ is the only minimizer of $\mathscr{E}$ (in the whole of $\mathscr{S}'$, if we extend it by assigning it the value $+\infty$ whenever $\alpha \in \mathscr{S}' \smallsetminus L^2_\varpi$), no convex combination of values of $\mathscr{E}$ other than $\mathscr{E}(-J/\varpi)$ itself can yield the ground state energy $E_0$. $\dashv$

The other equilibrium states of special interest we will focus upon are the so-called $\beta$-Gibbs states, where $\beta$ is the inverse temperature. Since $\Phi^0_t$ is the Hamiltonian flow of the quadratic Hamiltonian $\mathscr{E}^0 = \langle \alpha, \varpi \alpha \rangle_2$, its Gibbs measures are the Gaussian measures with covariance $(\beta\varpi)^{-1}$. The composition with the translation $\Theta_{\pm J/\varpi}$ yields a convolution with the measure $\delta_{-J/\varpi}$, that becomes a product in Fourier transform, yielding the following definition.

**Definition 2.11** (van Hove classical Gibbs states)**.**
For any source $J \in L^2_{1/\varpi}$, the *van Hove classical $\beta$-Gibbs states* $\omega^\beta_{0,\mathrm{Gibbs}} \in \mathrm{Reg}_0(\mathfrak{L}, \langle\,\cdot\,,\,\cdot\,\rangle_{\mathfrak{L}})_{+,1}$ are defined for any $\beta > 0$ by

$$\hat{\omega}^\beta_{0,\mathrm{Gibbs}}(f) = e^{-\frac{\pi^2}{\beta}\langle f, \varpi^{-1}f\rangle_2}e^{2\pi i \mathrm{Re}\langle f, -J/\varpi\rangle_2} \,.$$

Let us observe that for any $f \in \mathfrak{L}, f/\sqrt{\varpi} \in L^2$, and thus the Gibbs state is well defined. Furthermore, observe that as expected the ground state can be seen as the $\infty$-Gibbs state (zero temperature Gibbs state).

**Lemma 2.12.** *The classical $\beta$-Gibbs states $\omega^\beta_{0,\mathrm{Gibbs}}$ are $\beta(\mathscr{S}', \mathscr{S})$-Radon probability measures. Furthermore, they are all equilibrium states for $\tau_0$.*

*Proof.* It is not difficult to check that $\hat{\omega}^\beta_{0,\mathrm{Gibbs}} \in \mathfrak{G}(1)$; furthermore, $\hat{\omega}^\beta_{0,\mathrm{Gibbs}}\big|_{\mathscr{S}}$ is continuous on the whole of $\mathscr{S}$ and thus it defines a $\beta(\mathscr{S}', \mathscr{S})$-Radon probability measure by Bochner-Minlos theorem.

Concerning them being equilibria, by a direct computation:

$$\left(\tau_0(t)^{\mathrm{t}}\widehat{[\omega_{0,\mathrm{Gibbs}}^\beta]}\right)(f) = e^{-\frac{\pi^2}{\beta}\langle e^{it\varpi}f,\,\varpi^{-1}e^{it\varpi}f\rangle_2}\,e^{2\pi i\mathrm{Re}\langle e^{it\varpi}f,\,-\frac{J}{\varpi}\rangle_2}\,e^{2\pi i\mathrm{Re}\langle f,\,(e^{-it\varpi}-1)J/\varpi\rangle_2}$$

$$= e^{-\frac{\pi^2}{\beta}\langle f,\,\varpi^{-1}f\rangle_2}\,e^{2\pi i\mathrm{Re}\langle f,\,-\frac{J}{\varpi}\rangle_2} = \hat{\omega}_{0,\mathrm{Gibbs}}^\beta(f)\ ,$$

that together with Theorem 1.14 concludes the proof. ⊣

Let us conclude this section by discussing the long-time asymptotics, also called *scattering theory* for the classical van Hove model. We adopt an approach that is algebraic, and that is not customary in partial differential equations; nonetheless, it is very convenient to compare it with its quantum counterpart (defined in § 2.3.2, see § 3.3 for the comparison itself). Let us define the asymptotic characters $W_0^\pm(f)$ by comparing the van Hove dynamics with the source-less dynamics $\tau_0^0(t)$ whose action is

$$\tau_0^0(t)\big[W_0(f)\big] = W_0(e^{it\varpi}f)\ ,$$

for asymptotic long times $t \to \pm\infty$. This leads to the following definitions; beforehand, observe that $\tau_0^0(t) = \mathbb{W}_0(e^{it\varpi})$, if one sees $e^{it\varpi}$ as a group of symplectic automorphisms on $(\mathfrak{L}_\mathbb{R}, \mathrm{Im}\langle\,\cdot\,,\,\cdot\,\rangle_\mathfrak{L})$.

**Definition 2.13** (Scattering operators)**.**
Let $\tau$ be a dynamical map on $\mathbb{W}_0(\mathfrak{L}, \langle\cdot,\cdot\rangle_\mathfrak{L})$, seen as a perturbation of a reference "free map" $\mathbb{W}_0(s_t)$, $s_t : (\mathfrak{L}_\mathbb{R}, \mathrm{Im}\langle\cdot,\cdot\rangle_\mathfrak{L}) \to (\mathfrak{L}_\mathbb{R}, \mathrm{Im}\langle\cdot,\cdot\rangle_\mathfrak{L})$ a group of symplectic automorphisms. Then the *asymptotic characters* are defined as

$$W_0^\pm(f) = \lim_{t\to\pm\infty} \tau(t)\big[W_0(s_{-t}f)\big]\ ,$$

provided that the limits exist in the norm topology of $\mathbb{W}_0(\mathfrak{L}, \langle\cdot,\cdot\rangle_\mathfrak{L})$.

The *wave operators* $\Omega_0^\pm$ are the limit *-endomorphisms satisfying

$$W_0^\pm(f) = \Omega_0^\pm\big[W_0(f)\big]\ ,$$

and the *S-matrix* is the *-automorphism $S$ defined by

$$S_0 = \Omega_0^+ \circ \big(\Omega_0^-\big)^{-1}\ ,$$

provided $\Omega_0^\pm$ are invertible.

The physical interpretation of the wave operators is that they map (by transposition) regular states at finite time $\omega_0$ to asymptotic states $\omega_0^\pm = (\Omega_0^\pm)^{\mathrm{t}}\omega_0$, in such a way that the free evolution of the asymptotic state is close to the dynamical evolution of its image for asymptotic times: $\mathbb{W}_0(s_t)^{\mathrm{t}}\omega_0^\pm \sim_{t\to\pm\infty} \tau(t)^{\mathrm{t}}\omega_0$. The S-matrix connects asymptotic states $\omega_0^-$ in the past with asymptotic states $\omega_0^+$ in the future, however it requires the wave operators to be invertible, typically a rather strong property that takes the name of *asymptotic completeness*.

**Proposition 2.14.** *For any source $J \in L_{1/\varpi}^2$, the asymptotic characters for the van Hove model are given by*

$$W_0^\pm(f) = \Theta_{J/\varpi}\big[W_0(f)\big] = W_0(f)e^{2\pi i\mathrm{Re}\langle f, J/\varpi\rangle_2}\ .$$

*It then follows that*

$$\Omega_0^\pm = \Theta_{J/\varpi}\ ,$$

*and*

$$S_0 = \mathrm{id}\ .$$

*Furthermore, the collection of asymptotic abelian characters $\{W_0^\pm(f)\ ,\ f \in \mathfrak{L}\}$ is a unitary representation of the abelian group $(\mathfrak{L}, +)$ in any representation of $\mathbb{W}_0(\mathfrak{L}, \langle\,\cdot\,,\,\cdot\,\rangle_\mathfrak{L})$.*

*Proof.* A straightforward computation shows that

$$\tau_0(t)\big[W_0(e^{-it\omega}f)\big] = W_0(f)e^{2\pi i\mathrm{Re}\langle f,(1-e^{it\varpi})J/\varpi\rangle_2} = W_0(f)e^{2\pi i\mathrm{Re}\langle f,J/\varpi\rangle_2}e^{-2\pi i\mathrm{Re}\langle f,e^{it\varpi}J/\varpi\rangle_2} \ .$$

Now, $J/\sqrt{\varpi} \in L^2$, and $f \in \mathfrak{L}$, therefore, as a function of $|k|$, $\int_{\Sigma} f(|k|,\vartheta)J(|k|,\vartheta)|k|^{d-2}\mathrm{d}\vartheta$ is absolutely integrable (we denote by $\vartheta$ the angular coordinates in the polar decomposition of $k \in \mathbb{R}^d$) for any $d \geq 2$. By Riemann-Lebesgue lemma, it follows that $\lim_{|t|\to\infty}\langle f,e^{it\varpi}J/\varpi\rangle_2 = 0$; and therefore

$$W_0^{\pm}(f) = \lim_{t\to\pm\infty}\tau_0(t)\big[W_0(e^{-it\omega}f)\big] = W_0(f)e^{2\pi i\mathrm{Re}\langle f,J/\varpi\rangle_2} \ .$$

The identification of $\Omega_0^{\pm} = \Theta_{J/\varpi}$ is an immediate consequence, recalling that $W_0(f)[T] = e^{2\pi i\mathrm{Re}\langle f,T\rangle_2}$, and $\Theta$ is a translation in the phase space. Finally, it is not difficult to check that the collection of asymptotic characters $\{W_0^{\pm}(f) \ , f \in \mathfrak{L}\}$ satisfy the abstract properties of the abelian character group: for all $f,g \in \mathfrak{L}$,

- $W_0^{\pm}(f) \neq 0$,

- $W_0^{\pm}(f)^* = W_0^{\pm}(-f)$,

- $W_0^{\pm}(f)W_0^{\pm}(g) = W_0^{\pm}(f+g)$.

$\dashv$

### 2.3.2. *The quantum van Hove dynamical system.* To define the abstract van Hove dynamical system $\tau_\hbar$ on $\mathbb{W}_\hbar(\mathfrak{L},\langle\,\cdot\,,\,\cdot\,\rangle_{\mathfrak{L}})$, it is convenient to infer its properties from the map in Fock representation, and focus on the abstract definition mostly concerning the I-infrared singular case.

The quantum dynamical map can be inferred by the action of the strongly continuous unitary group $e^{-i\frac{t}{\hbar}H_\hbar}$ on the Fock space $\Gamma_{\mathrm{s}}(L^2(\mathbb{R}^d))$. It holds that

$$e^{i\frac{t}{\hbar}H_\hbar}W_\hbar(f)e^{-i\frac{t}{\hbar}H_\hbar} = W_\hbar(e^{it\varpi}f)e^{2\pi i\mathrm{Re}\langle f,(e^{-it\varpi}-1)J/\varpi\rangle_2} \ ,$$

that is perfectly analogous to the abelian case if one substitutes $W_0$ with $W_\hbar$. The map of unitary conjugation with $e^{i\frac{t}{\hbar}H_\hbar}$ extends to an isometric group of *-automorphisms on $\pi_{\mathrm{F}}\big(\mathbb{W}_\hbar(\mathfrak{L},\langle\,\cdot\,,\,\cdot\,\rangle_{\mathfrak{L}})\big)$, and thus it defines a dynamical map on the algebra itself.

**Definition 2.15** (Quantum van Hove dynamical map)**.**
For any source $J \in L^2_{1/\varpi}$, the isometric group of *-automorphisms $\{\tau_\hbar(t) \ , \ t \in \mathbb{R}\}$ on $\mathbb{W}_\hbar(\mathfrak{L},\langle\,\cdot\,,\,\cdot\,\rangle_{\mathfrak{L}})$ defined by

$$\tau_\hbar(t)\big[W_\hbar(f)\big] = W_\hbar(e^{it\varpi}f)e^{2\pi i\mathrm{Re}\langle f,(e^{-it\varpi}-1)J/\varpi\rangle_2} \ ,$$

is called the *quantum van Hove dynamical map*.

Its transposed action $\tau_\hbar(t)^{\mathrm{t}}$ on regular states $\omega_\hbar \in \mathrm{Reg}_\hbar(\mathfrak{L},\langle\,\cdot\,,\,\cdot\,\rangle_{\mathfrak{L}})_+$ is defined by

$$\big(\widehat{\tau_\hbar(t)^{\mathrm{t}}[\omega_\hbar]}\big)(f) = \hat{\omega}_\hbar(e^{it\varpi}f)e^{2\pi i\mathrm{Re}\langle f,(e^{-it\varpi}-1)J/\varpi\rangle_2} \ .$$

In the infrared regular case, the van Hove dynamical map has Fock-normal Gibbs states at any *quantum temperature*[6] $\beta_\hbar \leq \infty$. Using a very well-known C*-algebraic terminology[see, *e.g.*, BR97], we thus have Fock normal $(\beta_\hbar,\tau_\hbar)$-KMS states for any $\beta_\hbar \leq \infty$ for the infrared regular van Hove dynamical map.

**Lemma 2.16.** *For any $J \in L^2_{1/\varpi} \cap L^2_{1/\varpi^2}$, the coherent state $C_\hbar(-J/\varpi) \in \mathrm{Reg}_\hbar(\mathfrak{L},\langle\,\cdot\,,\,\cdot\,\rangle_{\mathfrak{L}})_{+,1}$ is Fock-normal, and it is the* unique *Fock-normal ground state of the quantum van Hove dynamical map $\tau_\hbar$.*

*Furthermore, the Gibbs states $\omega_{\hbar,\mathrm{Gibbs}}^{\beta_\hbar}$ induced by the Fock density matrices*

$$\varrho_{\hbar,\mathrm{Gibbs}}^{\beta_\hbar} = \frac{1}{\mathrm{Tr}e^{-\beta_\hbar H_\hbar}}e^{-\beta_\hbar H_\hbar}$$

---

[6]We discuss the dependence on $\hbar$ of the quantum temperature in detail below.

*are $(\beta_h, \tau_h)$-KMS states. Their noncommutative Fourier transform is given by*

$$\hat{\omega}_{h,\mathrm{Gibbs}}^{\beta_h}(f) = e^{-\frac{\pi^2 h}{2}\langle f, \coth(\beta_h \varpi/2)f\rangle_2} e^{2\pi i \mathrm{Re}\langle f, -J/\varpi\rangle_2}$$

*Proof.* The first result is a direct consequence of Theorem 2.6. Concerning Gibbs states, we rely on the well known and studied Gibbs states for the free Hamiltonian $\mathrm{d}\Gamma_h(\varpi)$. It is well known that for all $\varpi \geq 0$, $e^{-\beta_h \mathrm{d}\Gamma_h(\varpi)}$ is a positive trace class operator, yielding the free Gibbs state

$$\frac{1}{\mathrm{Tr}e^{-\beta_h \mathrm{d}\Gamma_h(\varpi)}}e^{-\beta_h \mathrm{d}\Gamma_h(\varpi)},$$

a $(\beta_h, e^{\frac{i}{h}\mathrm{d}\Gamma_h(\varpi)} \cdot e^{-\frac{i}{h}\mathrm{d}\Gamma_h(\varpi)})$-KMS state whose noncommutative Fourier transform is $e^{-\frac{\pi^2 h}{2}\langle f, \coth(\beta_h \varpi/2)f\rangle_2}$. The result then follows observing that $e^{-\beta_h H_h} = W_h(J/\pi i h \varpi)^* e^{-\beta_h(\mathrm{d}\Gamma_h(\varpi) - \|J\|^2_{L^2_{1/\varpi}})} W_h(J/\pi i h \varpi)$. ⊣

**Corollary 2.17.** *The van Hove ground state $C_h(-J/\varpi)$ is the weak-\* limit $\beta_h \to \infty$ of the $(\beta_h, \tau_h)$-Gibbs state $\omega_{h,\mathrm{Gibbs}}^{\beta_h}$.*

*Proof.* The pointwise limit of nonabelian Fourier transforms

$$\omega_{h,\mathrm{Gibbs}}^{\infty} := \lim_{\beta_h \to \infty} \hat{\omega}_{h,\mathrm{Gibbs}}^{\beta_h} = \hat{C}_h(-J/\varpi)$$

follows trivially from the properties of the hyperbolic cotangent and functional calculus for $\varpi$, seen as a multiplication operator in $L^2(\mathbb{R}^d)$. The convergence is then extended by linearity when testing any nonabelian trigonometric polynomial on the state, and by density to any element of $\mathbb{W}_h(\mathfrak{L}, \langle \cdot, \cdot \rangle_{\mathfrak{L}})$. ⊣

In order to define the equilibrium states in the I-infrared singular case, we shall use a non-Fock representation; we elucidate how these states emerge algebraically by taking the limit of regularized sources. Firstly, observe that the Gibbs states and the coherent state of Lemma 2.16, as regular states defined by their Fourier transform, make sense *for all $J \in L^2_{1/\varpi}$*. This leads to the following definition.

**Definition 2.18** (Algebraic Gibbs states). For any source $J \in L^2_{1/\varpi}$, let us define the $(\beta_h, \tau_h)$-Gibbs state, $\beta_h \leq \infty$, as the regular state $\omega_{h,\mathrm{Gibbs}}^{\beta_h} \in \mathrm{Reg}_h(\mathfrak{L}, \langle \cdot, \cdot \rangle_{\mathfrak{L}})_{+,1}$ defined by the Fourier transform

$$\hat{\omega}_{h,\mathrm{Gibbs}}^{\beta_h}(f) = e^{-\frac{\pi^2 h}{2}\langle f, \coth(\beta_h \varpi/2)f\rangle_2} e^{2\pi i \mathrm{Re}\langle f, -J/\varpi\rangle_2}.$$

**Proposition 2.19.** *For any $\beta_h < \infty$, $\omega_{h,\mathrm{Gibbs}}^{\beta_h}$ is a $(\beta_h, \tau_h)$-KMS state.*
*Furthermore, $\omega_{h,\mathrm{Gibbs}}^{\infty} = C_h(-J/\varpi)$ satisfies*

$$\omega_{h,\mathrm{Gibbs}}^{\infty} = \lim_{\beta_h \to \infty} \omega_{h,\mathrm{Gibbs}}^{\beta_h},$$

*where the limit is intended in the weak-\* topology.*

*Proof.* We use the following definition of the KMS-condition: $\omega_\beta \in \mathfrak{A}'_+$ is a $(\beta, \tau)$-KMS state if and only if for all $F \in \mathscr{F}^{-1}\mathscr{C}^{\infty}_c(\mathbb{R})$, and all $A, B \in \mathfrak{A}$,

$$\int_{\mathbb{R}} F(t - i\beta)\omega_\beta(A\tau(t)[B])\mathrm{d}t = \int_{\mathbb{R}} F(t)\omega_\beta(\tau(t)[B]A)\mathrm{d}t.$$

Firstly, let us prove this condition is satisfied by $\omega_{h,\mathrm{Gibbs}}^{\beta_h}$ when choosing $A = W_h(f)$ and $B = W_h(g)$. By an explicit computation, we have that

$$\omega_{h,\mathrm{Gibbs}}^{\beta_h}(W_h(f)\tau_h(t)[W_h(g)]) = e^{2\pi i \mathrm{Re}\langle g, (e^{-it\varpi}-1)J/\varpi\rangle_2 - i\pi^2 h \mathrm{Im}\langle f, e^{it\varpi}g\rangle_2} \omega_{h,\mathrm{Gibbs}}^{\beta_h}(W_h(f + e^{it\varpi}g))$$

$$= e^{2\pi i \mathrm{Re}\langle f+g, -J/\varpi\rangle_2 - \frac{\pi^2 h}{2}(\langle f, e^{it\varpi}g\rangle_2 - \langle g, e^{-it\varpi}f\rangle_2)}$$

$$e^{-\frac{\pi^2 h}{2}(\langle f, \coth(\beta_h \varpi/2)f\rangle_2 + \langle g, \coth(\beta_h \varpi/2)g\rangle_2 + \langle f, \coth(\beta_h \varpi/2)e^{it\varpi}g\rangle_2 + \langle g, \coth(\beta_h \varpi/2)e^{-it\varpi}f\rangle_2)}.$$

On the other hand,

$$\omega_{h,\text{Gibbs}}^{\beta_h}\big(\tau_h(t)[W_h(g)]W_h(f)\big) = e^{2\pi i \text{Re}\langle f+g, -J/\varpi\rangle_2 - \frac{\pi^2 h}{2}\big(\langle g, e^{-it\varpi}f\rangle_2 - \langle f, e^{it\varpi}g\rangle_2\big)}$$

$$e^{-\frac{\pi^2 h}{2}\big(\langle f, \coth(\beta_h\varpi/2)f\rangle_2 + \langle g, \coth(\beta_h\varpi/2)g\rangle_2 + \langle f, \coth(\beta_h\varpi/2)e^{it\varpi}g\rangle_2 + \langle g, \coth(\beta_h\varpi/2)e^{-it\varpi}f\rangle_2\big)} \ .$$

Therefore,

$$\int_{\mathbb{R}} F(t - i\beta_h)\,\omega_{h,\text{Gibbs}}^{\beta_h}\big(W_h(f)\tau_h(t)[W_h(g)]\big) = \int_{\mathbb{R}} F(t)\,\omega_{h,\text{Gibbs}}^{\beta_h}\big(W_h(f)\tau_h(t+i\beta_h)[W_h(g)]\big)\mathrm{d}t$$

$$= \int_{\mathbb{R}} F(t) e^{2\pi i \text{Re}\langle f+g, -J/\varpi\rangle_2 - \frac{\pi^2 h}{2}\big(\langle f, e^{it\varpi}e^{-\beta_h\varpi}g\rangle_2 - \langle g, e^{-it\varpi}e^{\beta_h\varpi}f\rangle_2\big)}$$

$$e^{-\frac{\pi^2 h}{2}\big(\langle f, \coth(\beta_h\varpi/2)f\rangle_2 + \langle g, \coth(\beta_h\varpi/2)g\rangle_2 + \langle f, \coth(\beta_h\varpi/2)e^{it\varpi}e^{-\beta_h\varpi}g\rangle_2 + \langle g, \coth(\beta_h\varpi/2)e^{-it\varpi}e^{\beta_h\varpi}f\rangle_2\big)}\mathrm{d}t$$

$$= \int_{\mathbb{R}} F(t)\,\omega_{h,\text{Gibbs}}^{\beta_h}\big(\tau_h(t)[W_h(g)]W_h(f)\big)\mathrm{d}t \ .$$

The result extends then to trigonometric polynomials by linearity, and to any almost periodic function by density.

The zero-temperature weak-* convergence of Gibbs states to $C_h(-J/\varpi)$ is proved as in Corollary 2.17.    ⊣

Let us now fix $J \in L^2_{1/\varpi} \smallsetminus L^2_{1/\varpi^2}$, and recall the definition of the infrared regularized source $J_n \in L^2_{1/\varpi} \cap L^2_{1/\varpi^2}$ as

$$J_n = \mathbb{1}_{|\cdot| \geq \frac{1}{n}} J \ .$$

The regularized source $J_n$ converges to $J$ in the weak-* topology of $\mathscr{S}'$. Let us now denote by $\tau_h^n$ the quantum van Hove dynamical system with source $J_n$, and by $\tau_h$ the quantum van Hove dynamical system with source $J$.

**Lemma 2.20.** *For any* $\beta_h \leq \infty$, *the* $(\beta_h, \tau_h^n)$-*Gibbs state* $\omega_{h,\text{Gibbs}}^{\beta_h, n}$ *of the regularized van Hove model converges, in weak-* topology as* $n \to \infty$, *to the* $(\beta_h, \tau_h)$-*Gibbs state* $\omega_{h,\text{Gibbs}}^{\beta_h}$ *of the I-infrared singular van Hove model.*

*Proof.* The proof of pointwise convergence of nonabelian Fourier transforms is straightforward, by definition of $J_n$. The convergence extends then to trigonometric polynomials by linearity and to all almost periodic functions by density.    ⊣

The above Lemma 2.20 is in the spirit of constructive quantum field theory: although algebraically we can define everything directly for any source $J \in L^2_{1/\varpi}$, it is customary to see a singular field theory as a suitable limit of its regularized version, in which a renormalization has been performed to "cure" the singularities. In the I-infrared singular van Hove model, the singularity amounts to the equilibrium states going out of the Fock representation, thus taking a representation-free approach is enough to "renormalize" this theory.

To conclude our algebraic treatment of equilibrium states, it remains to prove that $\omega_{h,\text{Gibbs}}^{\infty} = C_h(-J/\varpi)$ is a ground state for the quantum van Hove dynamical map in the I-infrared singular case $J \in L^2_{1/\varpi} \smallsetminus L^2_{1/\varpi^2}$.

**Proposition 2.21.** *For any* $J \in L^2_{1/\varpi}$, *the state* $\omega_{h,\text{Gibbs}}^{\infty} = C_h(-J/\varpi)$ *is an algebraic ground state for the van Hove dynamical map* $\tau_h$.

*Its GNS-representation is inequivalent to the Fock representation whenever* $J \in L^2_{1/\varpi} \smallsetminus L^2_{1/\varpi^2}$. *In that case, the GNS-Hilbert space is nonetheless*

$$\mathscr{H}_{\omega_{h,\text{Gibbs}}^{\infty}} = \Gamma_s(L^2(\mathbb{R}^d)) \ ,$$

*the van Hove dynamical map is implemented by the unitary operator* $e^{-i\frac{t}{h}\mathrm{d}\Gamma_h(\varpi)}$:

$$\pi_{\omega_{h,\text{Gibbs}}^{\infty}}\big(\tau_h(t)[A]\big) = e^{i\frac{t}{h}\mathrm{d}\Gamma_h(\varpi)}\pi_{\omega_{h,\text{Gibbs}}^{\infty}}(A)e^{-i\frac{t}{h}\mathrm{d}\Gamma_h(\varpi)} \ ;$$

*and* $\omega_{h,\text{Gibbs}}^{\infty}$ *acts as the projection on the Fock vacuum* $|\Omega_h\rangle\langle\Omega_h|$, *thus being the unique ground state of the van Hove Hamiltonian in this representation. Let us remark, however, that for all* $\mathfrak{L} \ni f \neq 0$, $\pi_{\omega_{h,\text{Gibbs}}^{\infty}}(W_h(f)) \neq e^{i\pi\varphi_h(f)}$, *this representation of* $\mathbb{W}_h(\mathfrak{L}, \langle\,\cdot\,,\,\cdot\,\rangle_{\mathfrak{L}})$ *being indeed non-Fock.*

*Proof.* An algebraic ground state $\omega^\infty$ for a dynamical map $\tau$ on a C*-algebra $\mathfrak{A}$ satisfies: for all $F \in \mathscr{F}^{-1}\mathscr{C}_c^\infty(\mathbb{R})$ with $\operatorname{supp}\hat{F} \subset \mathbb{R}_*^-$, and all $A, B \in \mathfrak{A}$:

$$\int_{\mathbb{R}} F(t)\,\omega^\infty\big(A\tau(t)[B]\big)\mathrm{d}t = 0 \;.$$

As shown in [ST71, Proposition 2.10], such property holds for any weak-* limit point as $\beta \to \infty$ of $(\beta, \tau)$-KMS states. Since by Proposition 2.19, $\omega_{h,\mathrm{Gibbs}}^\infty$ is the weak-* limit of $(\beta_h, \tau_h)$-KMS states, it is an algebraic ground state.

Its GNS-representation is studied in detail in [Ara20, §10.9.4], and it satisfies all the properties listed above in the I-infrared singular case[7]; the representation of the CCR is generated by a suitable, source-dependent, translation of the Fock-space Segal field $\varphi_h(\cdot)$. ⊣

Let us conclude this section again by studying the scattering theory for the van Hove dynamical map, in the quantum case. All ideas and results translate, *mutatis mutandis*, from the classical case.

**Definition 2.22** (Scattering operators)**.**
Let $\tau$ be a dynamical map on $\mathbb{W}_h(\mathfrak{L}, \langle \cdot, \cdot \rangle_{\mathfrak{L}})$, seen as a perturbation of a reference "free map" $\mathbb{W}_h(s_t)$, $s_t : (\mathfrak{L}_\mathbb{R}, \operatorname{Im}\langle \cdot, \cdot \rangle_{\mathfrak{L}}) \to (\mathfrak{L}_\mathbb{R}, \operatorname{Im}\langle \cdot, \cdot \rangle_{\mathfrak{L}})$ a group of symplectic automorphisms. Then the *asymptotic characters* are defined as

$$W_h^\pm(f) = \lim_{t \to \pm\infty} \tau(t)\big[W_h(s_{-t}f)\big] \;,$$

provided that the limits exist in the norm topology of $\mathbb{W}_h(\mathfrak{L}, \langle \cdot, \cdot \rangle_{\mathfrak{L}})$.

The *wave operators* $\Omega_h^\pm$ are the limit *-endomorphisms satisfying

$$W_h^\pm(f) = \Omega_h^\pm\big[W_h(f)\big] \;,$$

and the *S-matrix* is the *-automorphism $S_h$ defined by

$$S_h = \Omega_h^+\big(\Omega_h^-\big)^{-1} \;,$$

provided $\Omega_h^\pm$ are invertible.

Let us also define the *quantum translation operator* $\Theta_T^h$, $T \in \mathfrak{L}^\star$, to be the *-automorphism of $\mathbb{W}_h(\mathfrak{L}, \langle \cdot, \cdot \rangle_{\mathfrak{L}})$ defined by its action on the generators:

$$\Theta_T^h\big[W_h(f)\big] = W_h(f)e^{2\pi i \operatorname{Re}\langle f, T \rangle_2} \;.$$

**Proposition 2.23.** *For any source $J \in L_{1/\varpi}^2$, the asymptotic characters for the quantum van Hove model are given by*

$$W_h^\pm(f) = \Theta_{J/\varpi}^h\big[W_h(f)\big] = W_h(f)e^{2\pi i \operatorname{Re}\langle f, J/\varpi \rangle_2} \;,$$

*where $\Theta_{J/\varpi}^h$ is the quantum operator of translation. It then follows that*

$$\Omega_h^\pm = \Theta_{J/\varpi}^h \;,$$

*and*

$$S_h = \mathrm{id} \;.$$

*Furthermore, the collection of asymptotic nonabelian characters $\{W_h^\pm(f)\,,\ f \in \mathfrak{L}\}$ is a unitary representation of the Heisenberg group $\mathbb{H}_h(\mathfrak{L}_\mathbb{R}, \operatorname{Im}\langle \cdot, \cdot \rangle_{\mathfrak{L}})$ in any representation of $\mathbb{W}_h(\mathfrak{L}, \langle \cdot, \cdot \rangle_{\mathfrak{L}})$ (and generates a representation of the latter that is possibly inequivalent to the one of the original characters).*

*Proof.* The proof, apart from the last point, is completely analogous to the classical case, Proposition 2.14. Concerning the Heisenberg group property, it easy to check that the collection of asymptotic characters

---

[7]In the infrared regular case, its GNS-representation is unitarily equivalent to the Fock representation, the unitary map given by $W_h(J/\pi i h\varpi)$.

$\{W_{\hbar}^{\pm}(f)\ ,\ f\in\mathfrak{L}\}$ satisfy the abstract properties of the nonabelian Heisenberg character group: for all $f,g\in\mathfrak{L}$,

- $W_{\hbar}^{\pm}(f)\neq 0$,

- $W_{\hbar}^{\pm}(f)^* = W_{\hbar}^{\pm}(-f)$,

- $W_{\hbar}^{\pm}(f)W_{\hbar}^{\pm}(g) = W_{\hbar}^{\pm}(f+g)^{-i\pi^2\hbar\mathrm{im}\langle f,g\rangle_2}$.

The remark concerning inequivalent representations has the following example. Consider the I-infrared singular case in the Fock representation. The asymptotic Weyl operators can be computed explicitly, and they satisfy the properties above. However, the wave operators $\Omega_{\hbar}^{\pm}$ cannot be implemented as unitary operators in the Fock space, thus $C^*\{W_{\hbar}^{\pm}(f)\ ,\ f\in\mathfrak{L}\}$ is inequivalent to the Fock representation [see Der03, Theorem 4.2].                                                                                              ⊣

## 3. The correspondence principle for the van Hove model

In this section we study the correspondence principle $\hbar\to 0$ for the van Hove dynamical map quite extensively, with a focus on the dynamics itself (Egorov's theorem), the correspondence of equilibrium states, and of scattering. Let us remark that, compared to more realistic models of QFT for which the correspondence principle has been studied [see, *e.g.*, AB12, ABN19, AF14, AF17, AFH24, AFO23, AR20, AZ14, AKN01, ALN17, AJN15, AJN17, ALN16, AN16, CCFO21, CFO19, CF18, CFM23, CFO23a, CFO23b, Fal13], here the proofs are considerably less involved thanks to the fact that the van Hove model is explicitly solvable.

**3.1. Egorov's theorem.** Let us start with a dynamical version of the correspondence principle that in semiclassical analysis takes the name of *Egorov's theorem*. The idea is that the classical limit $\hbar\to 0$ and the evolution of regular states are "commuting operations". More precisely, Egorov's theorem expresses the commutativity of the following diagram:

$$
\begin{array}{ccc}
\omega_{\hbar_\beta} & \xmapsto{\ \tau_{\hbar_\beta}(t)^{\mathrm{t}}\ } & \omega_{\hbar_\beta}^t \\
{\scriptstyle\hbar_\beta\to 0}\Big\downarrow{\scriptstyle\mathscr{F}} & & {\scriptstyle\mathscr{F}}\Big\downarrow{\scriptstyle\hbar_\beta\to 0} \\
\omega_0 & \xmapsto{\ \tau_0(t)^{\mathrm{t}}\ } & \omega_0^t
\end{array}
$$

The precise result is stated in the theorem below. Beforehand, let us adopt the shorthand notation already used in the diagram above, namely

$$\omega_{\hbar}^t := \tau_{\hbar}(t)^{\mathrm{t}}[\omega_{\hbar}]\ ,$$
$$\omega_0^t := \tau_0(t)^{\mathrm{t}}[\omega_0]\ .$$

**Theorem 3.1** (Egorov)**.** *For any source* $J\in L_{1/\varpi}^2$, *the van Hove dynamical map satisfies a dynamical correspondence principle. More precisely, a family of regular quantum states* $\{\omega_{\hbar_\beta}\in\mathrm{Reg}_{\hbar_\beta}(\mathfrak{L},\langle\,\cdot\,,\cdot\,\rangle_{\mathfrak{L}})_{+,1}\ ,\ \hbar_\beta\to 0\}$ *satisfies* $\omega_{\hbar_\beta}\xrightarrow[\hbar_\beta\to 0]{\mathscr{F}}\omega_0\in\mathrm{Reg}_0(\mathfrak{L},\langle\,\cdot\,,\cdot\,\rangle_{\mathfrak{L}})_{+,1}$ *if and only if* $\omega_{\hbar_\beta}^t\xrightarrow[\hbar_\beta\to 0]{\mathscr{F}}\omega_0^t$ *for all* $t\in\mathbb{R}$.

*Proof.* The $\leftarrow$ implication is trivial (the convergence at $t=0$ in the hypothesis is the one to prove). To prove the $\to$ implication, observe that $\omega_{\hbar_\beta}\xrightarrow[\hbar_\beta\to 0]{\mathscr{F}}\omega_0$ means pointwise convergence of the Fourier transforms. On the other hand, for any $t\in\mathbb{R}$,

$$\hat{\omega}_{\hbar_\beta}^t(f) = (\tau_{\hbar_\beta}\widehat{(t)^{\mathrm{t}}[\omega}_{\hbar_\beta}])(f) = \hat{\omega}_{\hbar_\beta}(e^{it\varpi}f)e^{2\pi i\mathrm{Re}\langle f,(e^{-it\varpi}-1)J/\varpi\rangle_2}\ .$$

Therefore, we have that

$$\lim_{\hbar_\beta\to 0}\hat{\omega}_{\hbar_\beta}^t(f) = \hat{\omega}_0(e^{it\varpi}f)e^{2\pi i\mathrm{Re}\langle f,(e^{-it\varpi}-1)J/\varpi\rangle_2} = \hat{\omega}_0^t(f)\ .$$

⊣

### 3.2. The correspondence principle for equilibrium states.
We now turn to the study of $(\beta_\hbar, \tau_\hbar)$-Gibbs states ($\beta_\hbar \leq \infty$) in the limit $\hbar \to 0$. Let us make a couple of preliminary observations.

- As already noted, thanks to our abstract framework, we can consider on the same grounds all sources $J \in L^2_{1/\varpi}$ irrespective of the type of representation of the corresponding equilibrium states.

- Let us briefly and preliminarily discuss how the quantum temperature shall scale with $\hbar$. As we prove below, as long as there exists $0 < \varepsilon \leq 1$ such that $\beta_\hbar = O(\hbar^{1-\varepsilon})$, then the $(\beta_\hbar, \tau_\hbar)$-Gibbs state converges to the classical ground state, *i.e.* the $(\infty, \tau_0)$-Gibbs state. In other words, if the microscopic temperature remains finite in the limit $\hbar \to 0$ or diverges too slowly, the resulting classical temperature is zero. On the other hand, if there exists $\varepsilon > 0$ such that $\beta_\hbar = o(\hbar^{1+\varepsilon})$, then the limit is always singular. The semiclassical inverse temperature is thus $\beta_\hbar = \beta\hbar$, where $\beta \leq \infty$ is the macroscopic inverse temperature to which the $(\beta\hbar, \tau_\hbar)$-Gibbs state converges.

**Theorem 3.2** (Correspondence at equilibrium). *For any source $J \in L^2_{1/\varpi}$, for any $0 < \beta < \infty$, and for any $0 < \varepsilon \leq 1$, we have that:*

- $\omega^\infty_{\hbar,\mathrm{Gibbs}} \xrightarrow[\hbar \to 0]{\mathscr{F}} \omega^\infty_{0,\mathrm{Gibbs}}$

- $\omega^{\beta\hbar}_{\hbar,\mathrm{Gibbs}} \xrightarrow[\hbar \to 0]{\mathscr{F}} \omega^\beta_{0,\mathrm{Gibbs}}$

- $\omega^{O(\hbar^{1-\varepsilon})}_{\hbar,\mathrm{Gibbs}} \xrightarrow[\hbar \to 0]{\mathscr{F}} \omega^\infty_{0,\mathrm{Gibbs}}$

- $\widehat{\omega}^{O(\hbar^{1+\varepsilon})}_{\hbar,\mathrm{Gibbs}}(f) \xrightarrow[\hbar \to 0]{} \begin{cases} 1 & \text{if } f = 0 \\ 0 & \text{otherwise} \end{cases}$

*Proof.* The Fourier transform, and thus its convergence, is explicit for all the Gibbs states. More precisely:

- The Fourier transform of the quantum ground state $(C_\hbar(\widehat{-J/\varpi}))(f) = e^{-\frac{\pi^2\hbar}{2}\|f\|_2^2}e^{2\pi i\mathrm{Re}\langle f, -J/\varpi\rangle_2}$ converges to $\widehat{\delta}_{-J/\varpi}$, the Fourier transform of the classical ground state as already proved in Lemma 1.25;

- Using the Taylor expansion of the hyperbolic cotangent,

$$\lim_{\hbar \to 0} \widehat{\omega}^{\beta\hbar}_{\hbar,\mathrm{Gibbs}}(f) = \lim_{\hbar \to 0} e^{-\frac{\pi^2\hbar}{2}\langle f, \coth(\beta\hbar\varpi/2)f\rangle_2} e^{2\pi i\mathrm{Re}\langle f, -J/\varpi\rangle_2}$$
$$= e^{-\frac{\pi^2}{\beta}\langle f, \varpi^{-1}f\rangle_2} e^{2\pi i\mathrm{Re}\langle f, -J/\varpi\rangle_2} = \widehat{\omega}^\beta_{0,\mathrm{Gibbs}}(f) \ .$$

- If $\beta_\hbar = O(\hbar^{1-\varepsilon})$,

$$\lim_{\hbar \to 0} \widehat{\omega}^{O(\hbar^{1-\varepsilon})}_{\hbar,\mathrm{Gibbs}}(f) = \lim_{\hbar \to 0} e^{-\frac{\pi^2\hbar}{2}\langle f, \coth(O(\hbar^{1-\varepsilon})\varpi/2)f\rangle_2} e^{2\pi i\mathrm{Re}\langle f, -J/\varpi\rangle_2} = e^{2\pi i\mathrm{Re}\langle f, -J/\varpi\rangle_2} = \widehat{\delta}_{-J/\varpi}(f) \ .$$

- Finally, if $\beta_\hbar = O(\hbar^{1+\varepsilon})$, the limit is no more the Fourier transform of a regular classical state (it is a discontinuous function in zero):

$$\lim_{\hbar \to 0} \widehat{\omega}^{O(\hbar^{1+\varepsilon})}_{\hbar,\mathrm{Gibbs}}(f) = \lim_{\hbar \to 0} e^{-\frac{\pi^2\hbar}{2}\langle f, \coth(O(\hbar^{1+\varepsilon})\varpi/2)f\rangle_2} e^{2\pi i\mathrm{Re}\langle f, -J/\varpi\rangle_2} = \begin{cases} 1 & \text{if } f = 0 \\ 0 & \text{otherwise} \end{cases} \ .$$

⊣

### 3.3. The correspondence principle for asymptotic fields and scattering operators.

The final result we are able to prove − to showcase the robustness of our abstract framework in studying the correspondence principle − concerns long-time asymptotics and scattering operators. Given a regular state $\omega_\hbar \in \mathrm{Reg}_\hbar(\mathfrak{L}, \langle\,\cdot\,,\,\cdot\,\rangle_\mathfrak{L})$ ($\hbar \geq 0$), in light of Propositions 2.14 and 2.23 we can define its asymptotic Fourier transform

$$\hat{\omega}_\hbar^\pm(f) = \omega_\hbar\big(W_\hbar^\pm(f)\big)\ .$$

By Definitions 2.13 and 2.22, the asymptotic Fourier transform can be clearly seen also as the standard Fourier transform of the asymptotic state $\omega_\hbar^\pm = (\Omega_\hbar^\pm)^{\mathsf{t}}[\omega_\hbar]$:

$$\hat{\omega}_\hbar^\pm = \widehat{\omega_\hbar^\pm}\ .$$

Let us remark that, seen as a cylindrical measure, the asymptotic state $\omega_0^\pm$ is obtained by pushing forward the original state by means of the wave operator: $\omega_0^\pm = \Omega_0^\pm{}_* \omega_0$.

**Theorem 3.3** (Scattering correspondence). *For any source $J \in L^2_{1/\varpi}$, we have that any family of regular quantum states $\{\omega_{\hbar_\beta} \in \mathrm{Reg}_{\hbar_\beta}(\mathfrak{L}, \langle\,\cdot\,,\,\cdot\,\rangle_\mathfrak{L})_{+,1}\ ,\ \hbar_\beta \to 0\}$ satisfies $\omega_{\hbar_\beta} \xrightarrow[\hbar_\beta \to 0]{\mathscr{F}} \omega_0 \in \mathrm{Reg}_0(\mathfrak{L}, \langle\,\cdot\,,\,\cdot\,\rangle_\mathfrak{L})_{+,1}$ if and only if $\omega_{\hbar_\beta}^\pm \xrightarrow[\hbar_\beta \to 0]{\mathscr{F}} \omega_0^\pm$.*

*In other words, the following diagram is commutative:*

*Proof.* The implication → follows as usual by an explicit limit of the asymptotic Fourier transform:

$$\lim_{\hbar_\beta \to 0} \hat{\omega}_{\hbar_\beta}^\pm(f) = \lim_{\hbar_\beta \to 0} \hat{\omega}_{\hbar_\beta}(f) e^{2\pi i \mathrm{Re}(f J/\varpi)_2} = \omega_0(f) e^{2\pi i \mathrm{Re}(f J/\varpi)_2}$$

$$= \int_{\mathscr{F}'}^\bullet W_0(f)[T] e^{2\pi i \mathrm{Re}(f J/\varpi)_2} \mathrm{d}\omega_0(T) = \hat{\omega}_0^\pm(f)\ .$$

The converse implication ← is analogous, observing that the inverse wave operators are given by the reverse translation: given an asymptotic state $\omega_\hbar^\pm$ ($\hbar \geq 0$) with standard Fourier transform $\widehat{\omega_\hbar^\pm}$, it follows that the Fourier transform of the corresponding time-zero state $\omega_\hbar = (\Omega_\hbar^{-1})^{\mathsf{t}}\omega_\hbar^\pm$ is given by $\hat{\omega}_\hbar(f) = \widehat{\omega_\hbar^\pm}(f) e^{2\pi i \mathrm{Re}(f, -J/\varpi)_2}$. ⊣

Dipartimento di Matematica, Politecnico di Milano, P.zza Leonardo da Vinci 32, 20133 Milano
*Email address*: `marco.falconi@polimi.it`

Master Program in Physical Engineering, Politecnico di Milano, P.zza Leonardo da Vinci 32, 20133 Milano
*Email address*: `lorenzo1.fratini@mail.polimi.it`